\documentclass[]{emulateapj}
\usepackage{natbib}
\shorttitle{\sc A New Cepheid distance to NGC$\,$4258}
\shortauthors{\sc Macri et al.}
\newcommand{\nstt}{536}
\newcommand{\nret}{255}
\newcommand{\ntot}{281}

\newcommand{\mc}{\multicolumn{3}{l}}
\newcommand{\fracs}{{\mbox{\ensuremath{.\hspace*{-2pt}^{\rm s}}}}}
\newcommand{\nsth}{258}

\newcommand{\ntoh}{173}

\newcommand{\dmuo} {10.87\pm0.05_r\pm0.05_s}    
\newcommand{\noud} {20}
\newcommand{\dmui} {10.71\pm0.04_r\pm0.05_s}    
\newcommand{\nind} {69}

\newcommand{\mumsr}{29.29\pm0.09_r\pm0.12_s}
\newcommand{\dmu}  {10.88\pm0.04_r\pm0.05_s}    
\newcommand{\dmutip}{10.87\pm0.06_r\pm0.10_s}
\newcommand{\dmua} {$10.88\pm0.04$~(random) $\pm0.05$~(systematic)}    
\newcommand{\mulmc}{18.41\pm0.10_r\pm0.13_s}    
\newcommand{\dlmc} {48.1\pm2.3_r\pm2.9_s}       
\newcommand{\metz} {-0.29\pm0.09_r\pm0.05_s}    
\newcommand{\mett} {-0.49\pm0.15_r}             
\newcommand{\hnod} {74\pm3_r\pm6_s}
\newcommand{\muvkpo}{11.19\pm0.04_r\pm0.02_s}   
\newcommand{\muikpo}{11.05\pm0.03_r\pm0.02_s}   
\newcommand{\muokpo}{10.86\pm0.04_r\pm0.05_s}   
\newcommand{\nouk}  {38}
\newcommand{\muvkpi}{11.33\pm0.04_r}            
\newcommand{\muikpi}{11.07\pm0.03_r}            
\newcommand{\muokpi}{10.69\pm0.04_r\pm0.05_s}   
\newcommand{\nink}  {85}

\newcommand{\gal}{NGC$\,$4258}
\newcommand{\tuc}{NGC$\,$104}
\newcommand{\bvi}{\mbox{$BV\!I$}}
\newcommand{\ebv}{\mbox{$E(B\!-\!V)$}}
\newcommand{\evi}{\mbox{$E(V\!-\!I)$}}
\newcommand{\ebi}{\mbox{$E(B\!-\!I)$}}
\newcommand{\bi} {\mbox{$B\!-\!I$}}
\newcommand{\vi} {\mbox{$V\!-\!I$}}
\newcommand{\bvo}{\mbox{$(B\!-\!V)_0$}}
\newcommand{\vio}{\mbox{$(V\!-\!I)_0$}}
\newcommand{\bio}{\mbox{$(B\!-\!I)_0$}}
\newcommand{\nd}{\multicolumn{1}{c}{$\dots$}}
\newcommand{\ksm}{\ km s$^{-1}$ Mpc$^{-1}$}

\begin{document}
\title{A new Cepheid distance to the maser-host galaxy \gal\ \\ and its implications for the Hubble Constant\altaffilmark{1}}

\author{L.~M.~Macri\altaffilmark{2}}
\affil{National Optical Astronomy Observatory}
\affil{950 North Cherry Ave., Tucson, AZ 85719, USA}

\author{K.~Z.~Stanek}
\affil{Department of Astronomy, The Ohio State University}
\affil{140 West 18th Ave., Columbus, OH 43210, USA}

\author{D.~Bersier}
\affil{Astrophysics Research Institute, Liverpool John Moores University}
\affil{Twelve Quays House, Egerton Wharf, Birkenhead CH41 1LD, England}

\author{L.~J.~Greenhill}
\affil{Harvard-Smithsonian Center for Astrophysics}
\affil{60 Garden St., Cambridge, MA 02138, USA}

\and

\author{M.~J.~Reid}
\affil{Harvard-Smithsonian Center for Astrophysics}
\affil{60 Garden St., Cambridge, MA 02138, USA}

\altaffiltext{1}{Based on observations with the Advanced Camera for Surveys onboard the NASA/ESA Hubble Space Telescope, obtained at STScI, which is operated by AURA, Inc., under NASA contract NAS 5-26555. These observations are part of program \# GO-9810.}
\altaffiltext{2}{Hubble Fellow \& Goldberg Fellow}

\addtocounter{footnote}{2}

\begin{abstract}
We present initial results from a time-series BVI survey of two fields in \gal\ using the Advanced Camera for Surveys onboard the Hubble Space Telescope. This galaxy was selected because of its accurate maser-based distance, which is anticipated to have a total uncertainty of $\sim3\%$. The goal of the HST observations is to provide an absolute calibration of the Cepheid Distance Scale and to measure its dependence on chemical abundance (the so-called ''metallicity effect''). 

We carried out observations of two fields at different galactocentric distances with a mean abundance difference of 0.5~dex. We discovered a total of \ntot\ Cepheids with periods ranging from 4 to 45 days (the duration of our observing window). We determine a Cepheid distance modulus for \gal\ (relative to the LMC) of $\Delta\mu_0=$\dmua~mag. Given the published maser distance to the galaxy, this implies $\mu_0(LMC)=\mulmc$~mag or $D(LMC)=\dlmc$~kpc. We measure a metallicity effect of $\gamma=\metz$~mag~dex$^{-1}$. We see no evidence for a variation in the slope of the Period-Luminosity relation as a function of abundance. 

We estimate a Hubble Constant of $H_0=\hnod$\ksm\ using a recent sample of 4 well-observed type Ia SNe and our new calibration of the Cepheid Distance Scale. It may soon be possible to measure the value of $H_0$ with a total uncertainty of 5\%, with consequent improvement in the determination of the equation of state of dark energy.
\end{abstract}

\section{Introduction}
During the last 15 years, the {\em Hubble Space Telescope} ({\em HST}) has been used to discover $\sim 10^3$ Cepheid variables in $\sim 30$ galaxies with $D \lesssim 25\;$Mpc, mostly through $V$- and $I$-band observations carried out with the WFPC2 instrument. The distance moduli to these galaxies have been determined through the use of a fiducial Cepheid Period-Luminosity relation (P-L) based on observations of variables located in the Large Magellanic Cloud. Several secondary distance indicators (such as type Ia SNe, the Tully-Fisher relation, the Surface Brightness Fluctuation method) have been calibrated based on these Cepheid distances. As a result of these investigations, there is some agreement that H$_0$ is about 70\ksm, perhaps with as little as 10\% uncertainty \citep{freedman01}. However, two significant sources of systematic error stand out.

First, the entire Cepheid Distance Scale is underpinned by the distance to the Large Magellanic Cloud (LMC). The distance to that galaxy is used to establish the absolute calibration of the Cepheid P-L relations, and its uncertainty dominates the calibration of any secondary distance indicator. The suitability of the LMC for this purpose is problematic, since independent estimates of its distance disagree by as much as 0.5 mag, or 25\% \citep{benedict02}. Additionally, the internal structure of the galaxy along the line of sight remains poorly understood \citep{nikolaev04,vandermarel01}. Faced with this situation, most Cepheid-based determinations of $H_0$ have adopted $\mu_{LMC}=18.5\pm0.1$~mag, which corresponds to a distance of $D_{LMC}=50.1\pm2.3$~kpc.

Second, the effect of metal abundance on the Cepheid P-L relation is controversial. Several independent methods for an observational determination have yielded a variety of results \citep{sasselov97,kochanek97,kennicutt98,sakai04} with the {\it opposite sign} to what has been predicted by some theoretical investigations \citep{fiorentino02}, which also suggest a sensitivity to helium as well as metal content. Furthermore, the use of $V$ and $I$ photometry alone in previous HST surveys makes it difficult to disentangle the effects of reddening and metallicity and adds uncertainty to the determination of Cepheid distances.

We wish to establish a new Cepheid Distance Scale anchor galaxy, \gal, for which accurate geometric estimates of distance are available. \citet{herrnstein99} estimated its distance modulus to be $\mumsr$~mag, and it is anticipated that \citet{humphreys07} will reduce the total uncertainty of that estimate to $\la 3\%$. Our goal is even more compelling in light of the recent {\it WMAP} results \citep{spergel06} because many cosmological parameters depend sensitively on $H_0$ \citep[e.g.][]{eisenstein04,tegmark04,hu05}. An accurate geometric distance to \gal\ can also be used to directly calibrate secondary distance indicators, such as the tip of the red giant branch (TRGB).

This paper contains the first results of our project: deep time-series $\bvi$ photometry of two fields in \gal\ and the discovery and analysis of Cepheid variables. The paper is organized as follows: \S 2 contains details of the observations, data reduction and photometry, and the search for variables; \S 3 describes the selection criteria and the Cepheid samples; \S 4 presents the determination of a Cepheid distance to \gal, a measurement of the metallicity dependence of Cepheid-based distances, and a discussion of our results.

Throughout the paper, we denote random (statistical) uncertainties with a subscript {\it r} and systematic uncertainties with a subscript {\it s}, i.e., $\pm0.10_r\pm0.10_s$~mag.

\section{Observations, Data Reduction and Photometry}
\subsection{Observations}
We used the {\em Hubble Space Telescope} (GO program 9810) to observe two fields located at widely different galactocentric radii within the disk of \gal\ (M106). This spiral galaxy is one of the brightest members of the Coma-Sculptor Cloud \citep{tully87}, and has been classified as SAB(s)bc~II-III \citep{RC3} and Sb(s)~II \citep{RSA}.

We carried out the observations using the Advanced Camera for Surveys / Wide Field Camera \citep{ford03}, which consists of 2~back-illuminated SITe $2048\times 4096$ pixel CCDs. The average plate scale of the focal plane is $0\farcs05$ pix$^{-1}$, making each image $\sim 202\arcsec$ on the side. At the nominal \gal\ distance of $\sim 7.2\;$Mpc, this translates to a physical size of $\sim 1.7\;$pc pix$^{-1}$. 

\begin{figure}[ht]
\center{\includegraphics[width=6.5in]{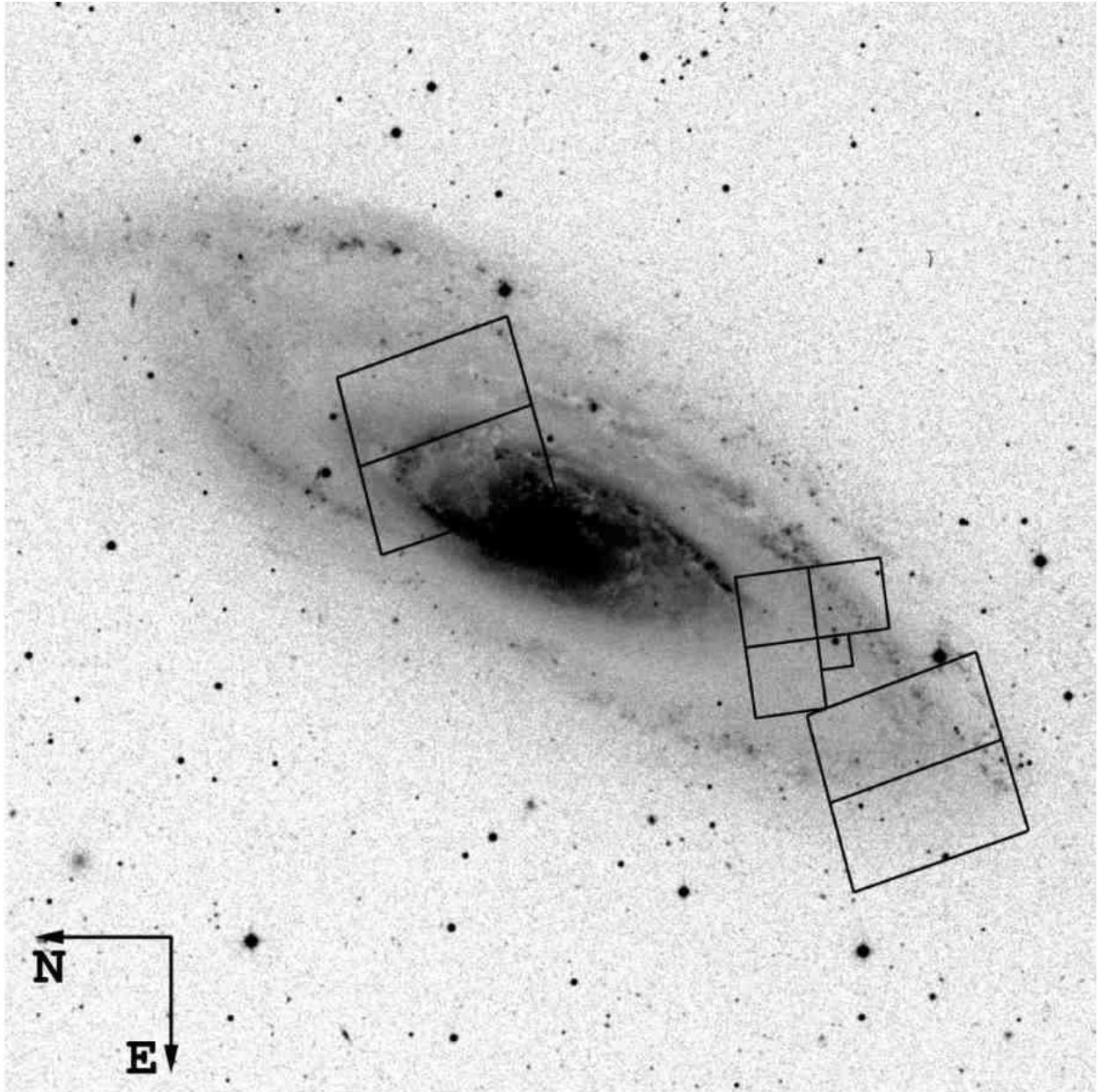}}
\caption{A blue DPOSS image of NGC$\,$4258 showing the two ACS/WFC fields observed for this project and the WFPC2 field previously studied by \citet{newman01}.}
\label{fig:fields}
\end{figure}

Hereafter, we refer to the two fields based on their galactocentric radii as ``inner'' and ``outer''. The fields are centered at $(\alpha,\delta)=$ $12^{\rm h}18^{\rm m}47.518^{\rm s}$, $+47\arcdeg20\arcmin20\farcs10$ (inner) and $12^{\rm h}19^{\rm m}23.891^{\rm s}$, $+47\arcdeg11\arcmin37\farcs61$ (outer), in J2000.0 coordinates. Figure~\ref{fig:fields} shows the location of these fields in the context of a digitized POSS-II image of \gal\footnote{The Second Palomar Observatory Sky Survey (POSS-II) was carried out by the California Institute of Technology with funds from NSF, NASA, the National Geographic Society, the Sloan Foundation, the Samuel Oschin Foundation, and the Eastman Kodak Corporation.  The Oschin Schmidt Telescope is operated by the California Institute of Technology and Palomar Observatory. The Digitized Sky Survey was produced at the Space Telescope Science Institute under U.S. Government grant NAG W-2166.}.

The fields were observed on twelve separate epochs between 2003 December 5 and 2004 January 19. The spacing of the visits followed a power-law distribution to minimize aliasing \citep{madore05}. Table~\ref{tab:log} contains a log of the observations. The fields were imaged in three colors on two consecutive orbits during each visit, following a standard two-point dither pattern that minimizes the effects of the geometric distortion present in ACS while ensuring a robust cosmic-ray rejection. Total exposure times per epoch were $2\times900s$ using the $F435W$ filter (similar to Johnson $B$), $2\times800s$ using the $F555W$ filter (similar to Johnson $V$) and $2\times400s$ using the $F814W$ filter (similar to Kron-Cousins $I$). One of the visits to the outer field was hampered by guide-star problems, reducing the total number of useful images by one relative to the inner field.

\subsection{Data Reduction and Photometry}
The raw observations were processed by the standard on-the-fly-reprocessing STScI ACS calibration pipeline, as described in the ACS Data Handbook \citep{pavlovsky05}. Briefly, the pipeline performs bias level correction and subtraction, dark image subtraction, flat fielding correction, and generation of ancillary data quality information. The calibrated images were downloaded from the STScI Archive and further processed using STSDAS and PyRAF\footnote{STSDAS and PyRAF are products of the Space Telescope Science Institute, which is operated by AURA for NASA}. Specifically, we used the PyDrizzle task to apply the filter-dependent geometric distortion correction to individual images.

We performed PSF photometry using the {\tt DAOPHOT/ALLSTAR/ALLFRAME} suite of programs \citep{stetson87,stetson94} and following the general data reduction and analysis precepts of the HST Key Project on the Extragalactic Distance Scale \citep[e.g.,][]{stetson98}. All the programs whose names appear in {\tt CAPITALS} were developed and kindly provided to us by Peter Stetson.

We defined the PSF as a quadratically-varying Moffat function with $\beta=1.5$ and a fitting radius of 2 pixels. The PSF extended out to a radius of 10 pixels ($0\farcs5$) and the local sky annulus was established from 20 to 25 pixels ($1-1\farcs25$). Aperture photometry was measured at logarithmically spaced radii from 3 to 10 pixels.

\subsubsection{Determination of template PSFs}
As expected in extragalactic Cepheid observations, our fields are rather crowded and lack bright, isolated stars suitable for the determination of the point-spread function (PSF). Given the long-term stability of {\it HST}, it is feasible to determine suitable PSFs from high S/N observations of dense yet uncrowded stellar fields. To this end, we retrieved a observations of an outer field of the globular cluster \tuc\ (program 9018), originally obtained for calibration purposes, from the HST Archive.

We analyzed 18 images in each of the $F435W$, $F555W$ and $F814W$ filters, taken at various offset positions. We used $\sim 800$ bright stars present in all the images to derive the PSF for each filter and to map its positional variation across the field of view of ACS/WFC. This was achieved using {\tt MULTIPSF}, which is identical to the {\tt PSF} routine in {\tt DAOPHOT} but uses stars in multiple images.

\subsubsection{PSF photometry}
\begin{figure}[ht]
\center{\includegraphics[width=6.5in]{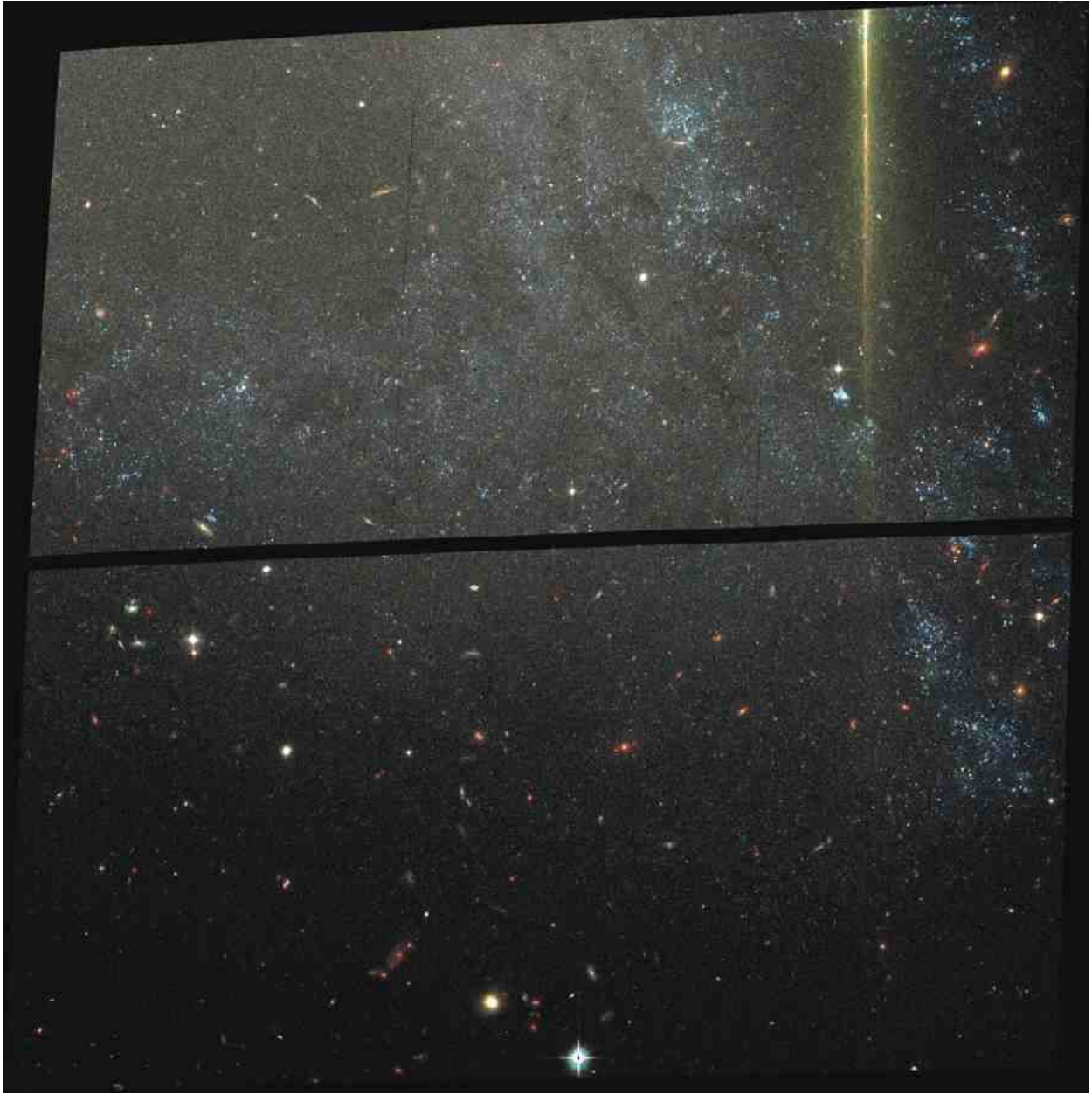}}
\caption{A color composite of the {\em HST} ACS/WFC outer field of \gal.}
\label{fig:ccou}
\end{figure}

\begin{figure}[ht]
\center{\includegraphics[width=6.5in]{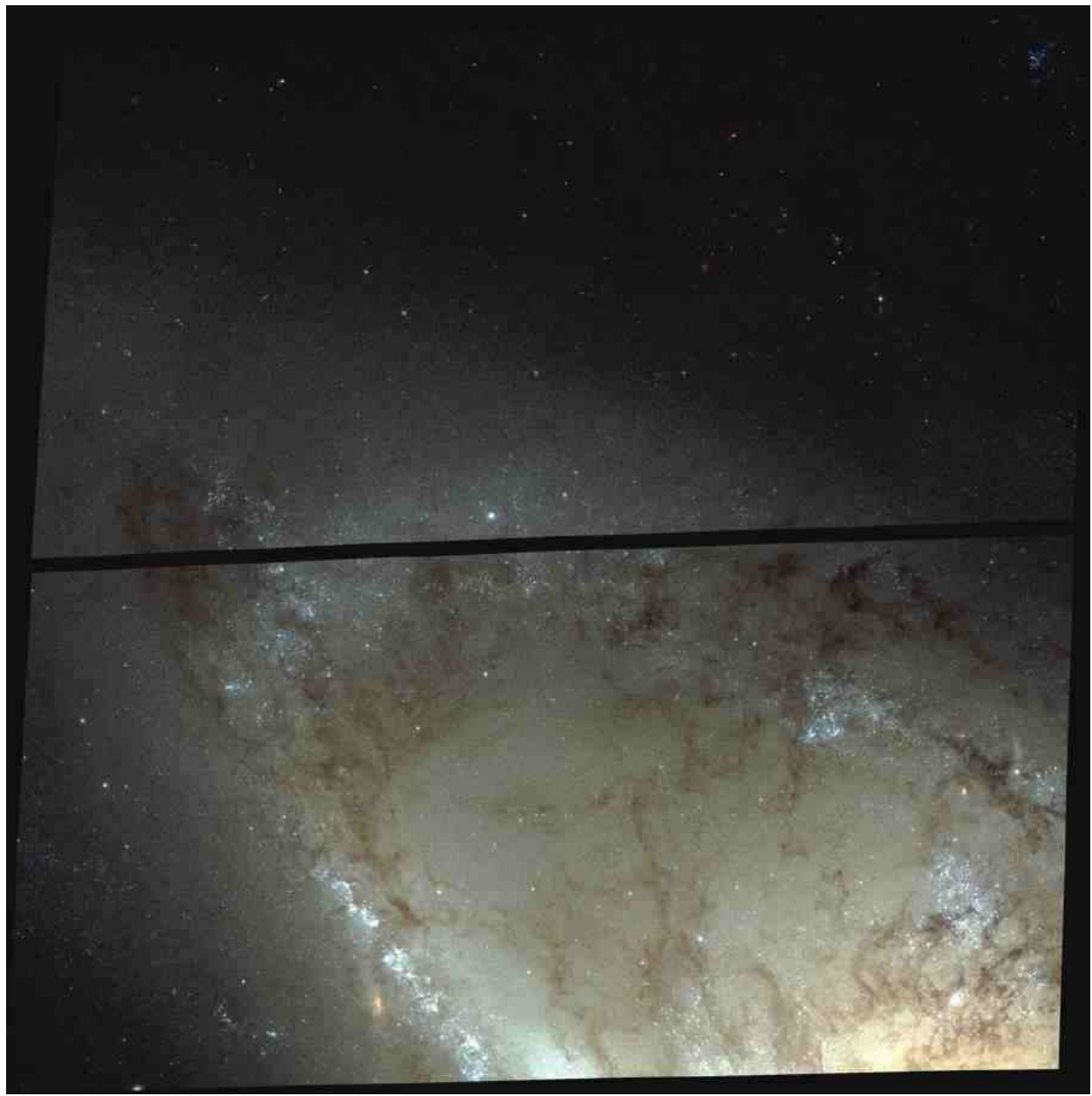}}
\caption{A color composite of the {\em HST} ACS/WFC inner field of \gal.}
\label{fig:ccin}
\end{figure}

We carried out PSF photometry separately for each combination of field and filter, as follows:

\noindent {\bf 1.~Image registration \& master image:} We used {\tt DAOPHOT} to detect bright stars in the individual images and {\tt ALLSTAR} to perform preliminary PSF photometry and obtain accurate coordinates of those objects. We used {\tt DAOMASTER} to determine coordinate transformations for every image, taking as reference the first image of each set. We used {\tt MONTAGE} to create a ``master'' image by interpolating the individual frames and applying a median filter. Figures~\ref{fig:ccou} \&~\ref{fig:ccin} are color composites of the ACS fields, created from the master $\bvi$ images.

\noindent {\bf 2.~Master object lists:} We detected objects in the master image following a two-step iterative approach (detect all objects, subtract them from the image, detect all remaining objects and add them to the initial list). At each iteration, we required a $3\sigma$ detection. The total number of objects detected were $\sim 2, 3$ and $6\times 10^5$ in $\bvi$, respectively, for the inner field and $\sim 0.6, 1$ and $2\times 10^5$ in $\bvi$, respectively, for the outer field.

\noindent {\bf 3.~PSF Photometry:} We used {\tt ALLFRAME} to measure the magnitudes of every star in each master list across all individual images in a given field/filter combination. We used the same template PSFs (\S 2.2.1) for all frames in a given filter. This generated a total of $\sim 2.3 \times 10^7$ photometric measurements.

\noindent {\bf 4.~Secondary standards:} We searched the star lists for bright, isolated stars suitable to serve as secondary standards. We identified $\sim 30-50$ suitable stars depending on the field and filter. We subtracted all other objects from each individual image and performed aperture photometry on these secondary standards to generate curves of growth. These were analyzed using {\tt DAOGROW} \citep{stetson90} and compared to the curves of growth determined from bright, isolated stars in \tuc. We found no difference between the two sets and decided to use both in our analysis.

\noindent {\bf 5.~Growth curves:} We used {\tt COLLECT} to apply the curves of growth and determine aperture corrections using the secondary standards. The corrections were small, with average values of $+0.02\pm0.04,-0.04\pm0.03,-0.05\pm0.02$~mag in $\bvi$, respectively. Epochs \#10 and \#11 had substantially larger aperture corrections ($\sim 0.2$~mag), probably due to telescope de-focusing.

\noindent {\bf 6.~Zeropoints:} We used {\tt CCDAVE} to compute mean instrumental magnitudes for the secondary standards. A typical r.m.s. scatter for these stars was $0.035$~mag, and as low as $0.015$~mag for the brightest objects. We used {\tt TRIAL} to compute the final zero-point corrections for each frame, using as a reference the mean aperture-corrected instrumental magnitudes of the secondary standards.

\noindent {\bf 7.~Astrometry:} We calculated celestial coordinates for all objects using the {\tt WCSTools/xy2sky} program \citep{mink02} and the astrometric solutions provided by STScI in the FITS headers of the first $F555W$ image of each field.

\subsection{Photometric calibration}
As a first step in our photometric calibration, we corrected the instrumental magnitudes for the effect of charge transfer efficiency (CTE) by applying Eqn.~(2) of the ACS Instrument Status Report 04-06 \citep{riess04}. We used the parameters derived by those authors for an aperture of 3 pixels in radius, since that is similar to the effective radius of the PSF for the bandpasses of interest (A. Riess, priv. comm.). Given the appreciable sky background in both inner and outer fields, the CTE correction was fairly small ($\sim 0.02$~mag).

We calibrated our photometry following the procedures of Appendix B of \citet{sirianni05}, using the zero-points and color terms listed under the ``observed'' columns of their Table 22:
\newcounter{subeqn}
\renewcommand{\theequation}{\arabic{equation}\alph{subeqn}}
\addtocounter{subeqn}{1}
\begin{equation}
V \, = \, F555W - AC05_V +25.704 - 0.054\, (V-I)\,\,\label{eqn:vvi}
\end{equation}
\addtocounter{equation}{-1}
\addtocounter{subeqn}{1}
\vspace*{-13pt}
\begin{equation}
V \, = \, F555W - AC05_V +25.701 - 0.056\, (B-V) \label{eqn:vbv}
\end{equation}
\renewcommand{\theequation}{\arabic{equation}}
\vspace*{-13pt}
\begin{equation}
I \,\, = \, F814W - \,AC05_I\,\, +25.495 - 0.002\, (V-I)\,\,\,\,\,\,
\end{equation}
\vspace*{-13pt}
\begin{equation}
B \,\, = \, F435W - AC05_B +25.842 - 0.089\, (B-V) \hspace{6pt}
\end{equation}

\noindent where $\bvi$ are the standard magnitudes and $F435W$, $F555W$, $F814W$ are the CTE-corrected, aperture-corrected (to $0\farcs5$) instrumental magnitudes derived in \S 2.2.2. The additional filter-dependent aperture corrections to infinity, $AC05_i$, are listed in Table 5 of \citet{sirianni05}. In the case of objects with three-color photometry (such as all the Cepheid variables), we gave preference to (\ref{eqn:vvi}) over (\ref{eqn:vbv}) because our $I$-band observations have higher $S/N$ than our $B$-band observations. Since these transformation equations make use of the standard (rather than observed) colors, they were applied iteratively until convergence.

Table~\ref{tab:secstd} lists the positions and calibrated magnitudes of the secondary standards to facilitate future comparisons with our work.

\subsection{Search for Variables and Classification}
\begin{figure}[ht]
\center{\includegraphics[angle=-90,scale=0.8]{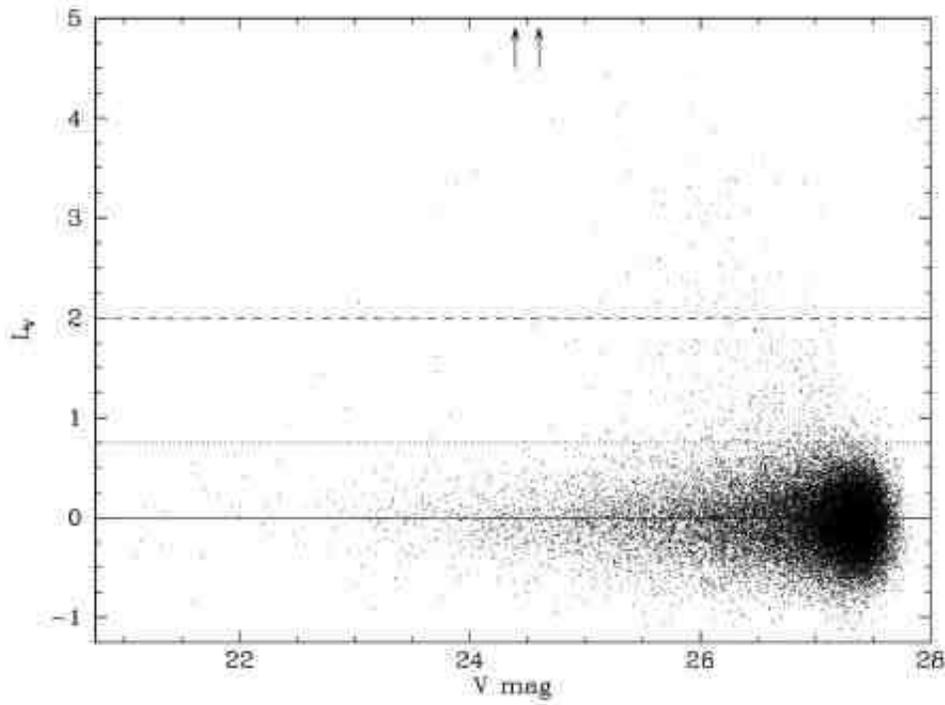}}
\caption{Distribution of the modified Welch-Stetson variability index $L_V$ \citep{stetson96} as a function of $V$ magnitude for objects in the outer field, after sigma-scaling. The dotted and dashed lines represent the minimum values of $L_V$ for the extended and restricted samples, respectively. Two variables with $L_V>6$ are represented by arrows.}
\label{fig:varidx}
\end{figure}

We searched for variables using the {\tt TRIAL} program, which performs a scaling of the reported {\tt ALLFRAME} measurement errors and calculates robust mean magnitudes and modified Welch-Stetson variability indices $L_V$ \citep{stetson96}. Figure~\ref{fig:varidx} shows the distribution of $L_V$ as a function of $V$ magnitude for the outer field (the inner field distribution is very similar but denser). By construction, the mean value of $L_V$ is zero. Given the observed $1\sigma$ dispersion in $L_V$ of 0.25, we set $L>0.75$ as the minimum variability threshold. We calculated the twenty most likely periods for each variable using the Lafler-Kinman algorithm \citep{lafler65} as encoded in {\tt TRIAL}.

We applied an automated classification algorithm developed by the DIRECT project \citep{kaluzny98} to the {\it V}-band light curves of the variables. The algorithm computed the chi-squared per degree of freedom, $\chi^2_\nu$, of each light curve for three cases: i) a constant magnitude (null hypothesis); ii) a linearly-varying magnitude (appropriate for objects with periods much longer than our observing window); iii) a Cepheid variable with a period equal to each one of the twenty tentative periods returned by the Lafler-Kinman algorithm. The latter case used the Cepheid template light curves developed by \citet{stetson96}.

We selected as possible Cepheids those periodic variables with a $\chi^2_\nu$ for case (iii) that was at least a factor of two smaller than the $\chi^2_\nu$ of cases (i) or (ii), following the methodology of the DIRECT project. Adopting the best-fit period, we phased the {\it B}- and {\it I}-band light curves and fit them with Cepheid template light curves (absent {\it B}-band templates, we used suitably scaled {\it V}-band ones). Finally we phase-weighted mean magnitudes \citep{stetson96} through numerical integration of the best-fit template light curve for each variable.

\section{Selection of Cepheid Variables}
\subsection{Adopted Period-Luminosity relations}
Our analysis requires the adoption of fiducial Period-Luminosity (P-L) relations to calculate distance moduli, as well as corresponding Period-Color (P-C) relations to correct for the effects of interstellar extinction. We adopted the P-L relations originally derived by \citet{udalski99}, as updated in the OGLE web site\footnote{\url{ftp://sirius.astrouw.edu.pl/ogle/ogle2/var\_stars/lmc/\hfill\\ cep/catalog/README.PL}}. These relations are based on a sample of $N>600$ Cepheids observed as part of the OGLE II project, with periods ranging from 2 to 30 days:
\begin{eqnarray}
B & = & 14.929(31) - 2.439(46)\ [\log P - 1] \label{eqn:plob}\\
V & = & 14.287(21) - 2.779(31)\ [\log P - 1] \label{eqn:plov}\\
I & = & 13.615(14) - 2.979(21)\ [\log P - 1] \label{eqn:ploi}
\end{eqnarray}

\noindent where $P$ is the period of the Cepheid in days, and the errors in the zero-points and slopes are expressed in units of $10^{-3}$. The dispersions of the data relative to the relations are 0.24, 0.16 and 0.11~mag in $\bvi$, respectively.

\subsection{Extinction corrections and relative distance moduli}
\gal\ is located at $l=138\arcdeg\!\!.32$, $b=68\arcdeg\!\!.84$. We estimated the value of foreground Galactic interstellar extinction based on the values in the reddening map of \citet{schlegel98} for a number of positions near the galaxy.  All of them yielded very low values of foreground extinction, $E(B-V)=0.016$~mag. We expect little additional (internal) extinction in the outer field, but the Cepheids in the inner field should be subject to a considerably larger amount of internal extinction with strong variations as a function of position.

We determined the total extinction to each Cepheid by comparing the observed $\bv$, $\vi$ and $\bi$ colors with the zero-extinction colors $\bvo$, $\vio$ and $\bio$ predicted by the P-C relations formed by Eqns.~(\ref{eqn:plob}-\ref{eqn:ploi}). We transformed the values of $\evi$ and $\ebi$ to $\ebv$ using the values of $A_\lambda$ from Table~6 of \citet{schlegel98} for $R_V=3.1$ and the extinction law of \citet{cardelli89}. We used the three color excesses to compute a mean $\ebv$ and standard deviation, but adopted a 0.025~mag uncertainty floor to account for the intrinsic width of the P-C relations.

We determined extinction-corrected LMC-relative distance moduli for each Cepheid by calculating:
\begin{eqnarray}
\Delta\mu_0 & = & \Delta\mu_I - 1.45 \evi \label{eqn:muovi}\\
\Delta\mu_0 & = & \Delta\mu_I - 2.38 \ebi \label{eqn:muobi}\\
\Delta\mu_0 & = & \Delta\mu_I - 1.94 \ebv \label{eqn:muobv}
\end{eqnarray}

\noindent{where $\Delta\mu_I$ is obtained by subtracting the mean value of $I(P)$ from Eq.~(\ref{eqn:ploi}) from the mean $I$-band magnitude of a given Cepheid of period $P$. The values of total-to-selective extinction ratios were calculated using the $A_\lambda/\ebv$ values in Table~6 of \citet{schlegel98}. The three values of $\Delta\mu_0$ were averaged to obtain a mean value and standard deviation. Eqn.~(\ref{eqn:muovi}) is the classical Wesenheit distance modulus used by \citet{freedman01} and \citet{saha01a}, although these authors apply it to distance moduli determined from an ensemble of Cepheids. We chose to form the additional two relations (Eqns.~\ref{eqn:muobi}-\ref{eqn:muobv}) to rule out any global systematic error in the calibration of the photometry.}

We note that in this analysis, all the calculated distance moduli are {\it relative} to the LMC, since the adopted P-L relations are expressed in {\it observed} magnitudes. We adopt this approach to accommodate the anticipated improvement in the geometric distance estimate for \gal\ to be obtained from new analyses of the maser observations \citep{humphreys07}.

\subsection{Selection criteria}
The selection of Cepheids from a larger set of periodic variables is not a trivial undertaking for {\em HST} observations, especially given the crowded nature of the fields, the sparse sampling of the light curves and the relatively low $S/N$ of the individual data points at the faintest magnitudes (corresponding to the shortest periods). Different studies have adopted various selection techniques, some based on visual inspections of light curves and images \citep{saha96} and some based on a more mathematical approach \citep{leonard03}.

In the case of these observations of \gal\ ---which is located significantly closer than most Cepheid-bearing galaxies studied with {\it HST}--- it is fairly easy to select a sample of high-quality Cepheids with $P\gtrsim 10$~days for the primary scientific goals of deriving a distance and measuring the metallicity effect. The values we obtain are insensitive to the application of different selection techniques. The selection of shorter-period ($P\lesssim 10$~d) Cepheids is less certain, especially in the case of the inner field. We have adopted a particular set of selection criteria, but we list all detected Cepheid candidates to facilitate alternative analyses by others.

We restricted the sample of variables using two cuts in $L_V$: a fairly low value of 0.75 (hereafter, the ``extended sample'') and a more conservative value of $L_V=2$ (hereafter, the ``restricted sample''). We applied additional selection criteria based on observed properties (light curve amplitude ratios, colors, etc.) in an attempt to remove contaminated Cepheids from the sample.

We detected a total of $\sim 10^6$ distinct objects at the $>3\sigma$ level in the master images. To exclude false positives, we trimmed objects detected in less than 75\% of the individual images of each band.

As stated in \S2.4, variable stars were classified as Cepheids if the $\chi^2_\nu$ was reduced by more than a factor of two (relative to the null hypothesis) when fitting the phased $V$-band data with a template Cepheid light curve. These requirements were met by $\nstt$ objects in the extended sample and $\nsth$ objects in the restricted sample. We applied the following secondary selection criteria:

\noindent {\bf 1.~Amplitude ratios:} Relative $\bvi$ amplitudes for fundamental-mode pulsators obey the proportions $1.5\!:\!1\!:\!0.5$. To discard obvious blue or red blends, we required {\bf (a)} an $I$-to-$V$ amplitude ratio of $0.5\pm0.25$ and {\bf (b)} a $B$-to-$V$ amplitude ratio of $1.5\pm0.5$.

\noindent {\bf 2.~Blue edge:} We discarded objects with $\ebv$ more than $2\sigma$ below the Galactic foreground value of 0.016~mag, as these variables are likely to be blended with blue stars. We set $\ebv=0.016$~mag for objects between the threshold and the foreground value.

\noindent {\bf 3.~Large extinction:} We rejected objects with $\ebv>0.5$~mag. These Cepheids are either blended with red stars or highly reddened, in which case the actual value of $R_V$ could deviate significantly from the adopted value of 3.1.

\noindent {\bf 4.~Pop II Cepheids:} We imposed a conservative upper limit of $\Delta\mu_0 < 12$~mag to the LMC-relative distance moduli to reject long-period RV Tauri and W Virginis variables in our sample. These Population II Cepheid-like variables obey period-color relations that are similar to those of Cepheids, and therefore had passed the previous selection criteria. However, their absolute magnitudes are more than 1~mag fainter than Cepheids.

Once these cuts were applied, we computed a mean modulus for each sample using a least-absolute-deviation technique with iterative sigma clipping. This was motivated by the asymmetric tails of outliers that are caused by other sources of contamination, many of which make objects artificially brighter.

\subsection{Final Cepheid samples}

\begin{figure}[ht]
\center{\includegraphics[angle=-90,scale=0.8]{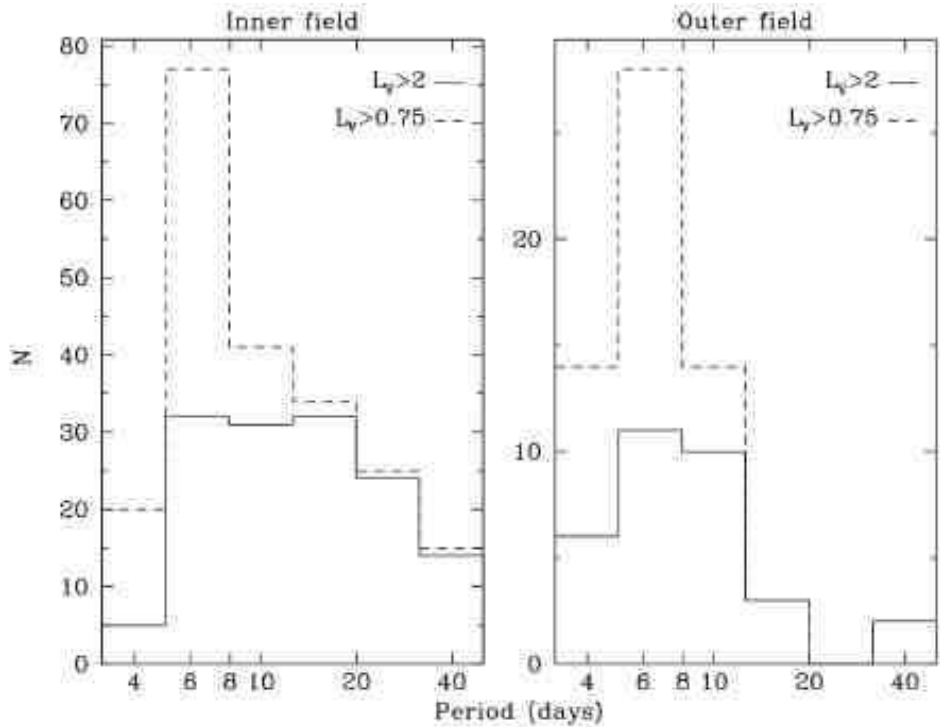}}
\caption{Period histograms for the final Cepheid samples after application of the selection criteria listed in \S3.3. Left panel: Inner field. Right panel: Outer field. Dashed line: extended sample. Solid line: restricted sample.}
\label{fig:phist}
\end{figure}

Table~\ref{tab:sel} details the effects of the selection criteria (\S3.3) on the initial samples. $\ntot$ out of $\nstt$ variables in the extended sample pass all criteria, while $\ntoh$ out of $\nsth$ variables in the restricted sample remain. Figure~\ref{fig:phist} shows the period distributions for both cuts in $L_V$. We used the restricted sample ($L_V>2$) in the subsequent analysis.

The observed properties of the $\ntot$~Cepheids that passed our selection criteria are listed in Table~\ref{tab:cephb}, while those of the $\nret$~rejected candidates are listed in Table~\ref{tab:rej} along with the reasons for their rejection. Table~\ref{tab:cephd} lists the derived properties of the Cepheids in Table~\ref{tab:cephb}. Table~\ref{tab:cepphot} contains the individual photometric measurements of these objects.

\begin{figure}[ht]
\center{\includegraphics[width=6.5in]{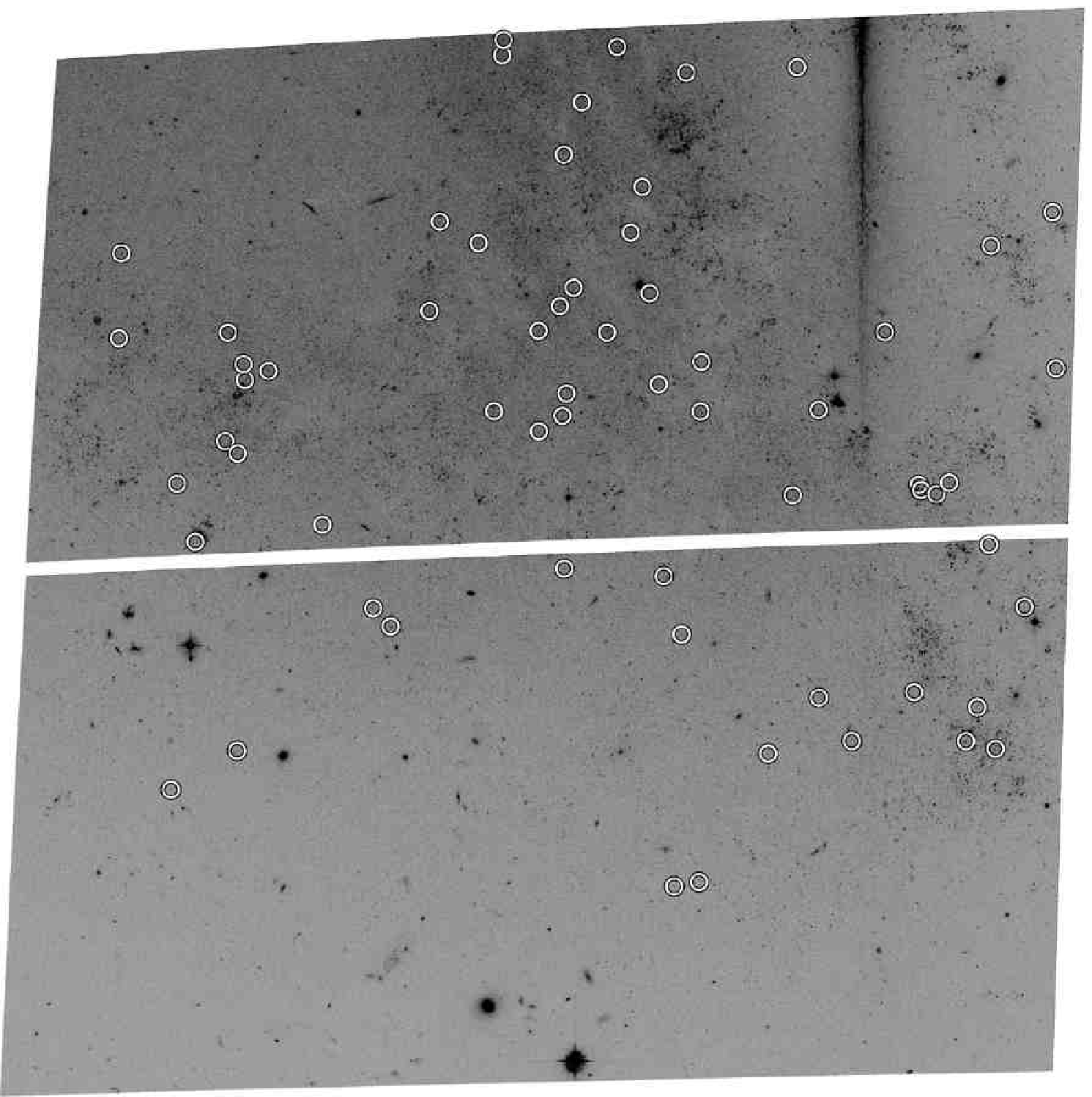}}
\caption{Master $V$-band image of the ACS outer field. The location of the Cepheids listed in Table~\ref{tab:cephb} are marked with open circles.}
\label{fig:fldou}
\end{figure}

\begin{figure}[ht]
\center{\includegraphics[width=6.5in]{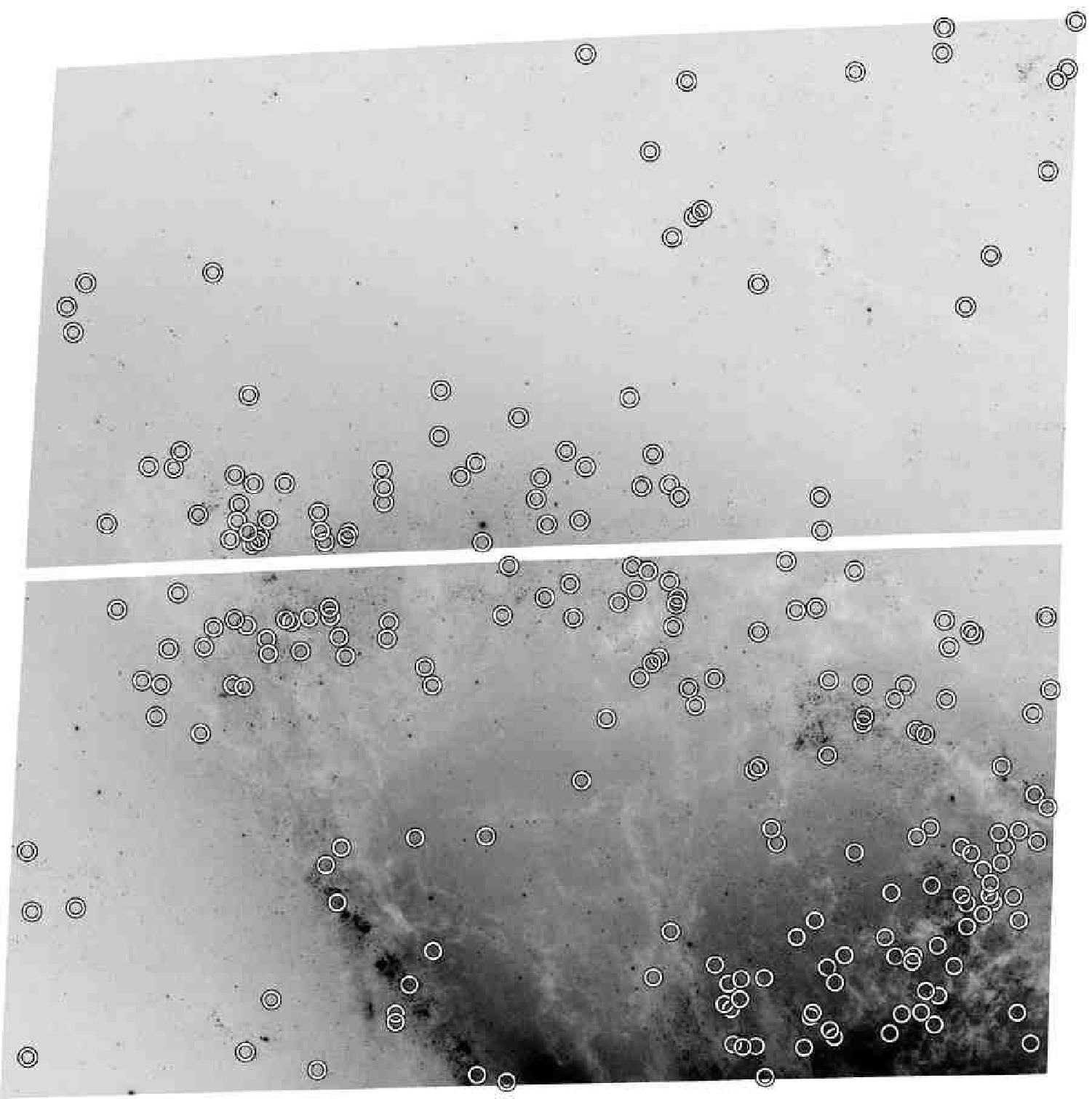}}
\caption{Master $V$-band image of the ACS inner field. The location of the Cepheids listed in Table~\ref{tab:cephb} are marked with open circles.}
\label{fig:fldin}
\end{figure}

Figures~\ref{fig:fldou} \&~\ref{fig:fldin} show the distribution of the Cepheids within the outer and inner fields, respectively; individual finding charts can be see in Figures~\ref{fig:fchart}-g. Figures~\ref{fig:cmdou} \&~\ref{fig:cmdin} show the distribution of the Cepheids within the color-magnitude diagrams of the two fields. Representative light curves are shown in Figures~\ref{fig:lcou} \&~\ref{fig:lcin}. Lastly, Figures~\ref{fig:plou} and \ref{fig:plin} contain the observed $\bvi$ P-L relations of the restricted samples.

\section{Results and Discussion}
\subsection{The Maser Distance to \gal}
Water maser emission observed from \gal\ originates in a subparsec annular region within a nearly edge-on, warped accretion disk, bound by a supermassive black hole in the nucleus \citep{miyoshi95,greenhill95}. Masers lie: (1) in a narrow sector on the near side of the disk and (2) on the disk-diameter perpendicular to the line of sight.

Geometric estimates of distance may be obtained from measurements of the centripetal acceleration or the proper motion of masers on the near side of the disk. The acceleration is obtained from the time rate of change of the maser Doppler shifts, and the proper motion is obtained from the change in the positions of the near-side masers relative to the approaching/receding masers (which appear to be stationary on the sky).

\citet{herrnstein99} reported acceleration and proper motion distance moduli that agreed to $<1\%$: $\mu_{\rm maser}=\mumsr$~mag. The quoted systematic uncertainty arises largely from unmodeled structure and an upper limit on the eccentricity of the disk. Initial models assumed circular orbits and a warp in position angle alone. More recently, \citet{herrnstein05} performed a detailed analysis of the disk rotation curve and detected a $2\sigma$ deviation from a Keplerian law, which they attributed to an inclination-warp in the disk. That also helps to explain the locus of the near-side masers.

\citet{humphreys07} aim to reduce the random component of the uncertainty by including more epochs of observation, and more importantly, to reduce the systematic component by improving the dynamical model of the maser-disk system. The \citet{herrnstein99} distance relied on VLBI data collected at four epochs between 1994 and 1997, while data for 18 VLBI epochs (1997-2000) and 40 spectroscopic epochs (1994-2003) are now available. The analysis also limited disk eccentricity to $\la 0.1$. More densely sampled data with a longer time baseline, coupled with a more sophisticated model of the disk warp and eccentricity, are anticipated to reduce the systematic and random uncertainties in distance by more than a factor of two, for a total uncertainty of $\sim 3$\% \citep{humphreys05b,humphreys05a}.

\subsection{A Cepheid distance to \gal}

\subsubsection{Minimum period cut}

We imposed minimum period cuts to the samples derived in \S3.4 before we determined mean relative distance moduli. Several reasons motivate the use of such a cut.

\noindent {\it a)}~We are unable to differentiate between fundamental and overtone pulsators due to our sparse phase sampling. Overtone pulsators in the Magellanic Clouds have $2<P<6$~days and are $\sim 0.75$~mag brighter than fundamental pulsators with the same period \citep{udalski99}. Hence, they can produce a large systematic bias in the derived distance. 

\noindent {\it b)}~Confusion noise introduces a systematic bias in the photometry of Cepheids that becomes increasingly important at faint magnitudes, especially in the $I$-band \citep{saha90,saha96}.

\noindent {\it c)}~Observing objects near the detection limit may result in incompleteness bias at the shortest periods of the observed P-L relation \citep{sandage88}.

\noindent {\it d)}~The observed magnitudes of short-period Cepheids are more likely to be contaminated by unresolved blends with other disk stars \citep{mochejska00}, especially in the denser regions of the inner field.

We applied the cut at minimum period and calculated the mean value of the individual relative distance moduli following the procedure described in \S3.3. Figures~\ref{fig:dmou} \&~\ref{fig:dmin} show the impact of this procedure for the outer and inner fields. Figure~\ref{fig:dmp} shows the mean relative distance modulus and its uncertainty as a function of $P_{min}$ for both fields.

There is no statistically-significant variation in the mean relative distance modulus of the outer field as a function of $P_{min}$. The primary use of the outer field Cepheids in this study is to test the maser distance of \gal\ against the distance to the LMC without having to worry about abundance differences (since they have the same mean metallicity). We chose $P_{min}=6$~d as the final period cut for this sample to avoid contamination by overtone pulsators and to maximize the sample size and the overlap of period ranges between these two galaxies; note that $P_{max}$ is 32~d for the OGLE LMC sample and 44~d for \gal.

The inner field exhibits a mild trend with shorter distance moduli for smaller minimum period cut-offs, with a statistical significance of $\sim2.7\sigma$ ($P_{min}$ of 20~d vs. 5~d). We chose $P_{min}=12$~d to avoid the observed bias in distance modulus at shorter periods. This value of $P_{min}$ is similar to the typical lower limit of the Cepheid samples discovered in other galaxies observed with {\it HST} ($P_{min}=10-15$~d).

\subsubsection{Distance moduli}

Taking the aforementioned period cuts into account, and using the restricted samples, we derive distance moduli {\it relative to the LMC} of $\Delta\mu_0=\dmuo$~mag (outer field, $N=\noud$ Cepheids) and $\Delta\mu_0=\dmui$~mag (inner field, $N=\nind$ Cepheids). 

The quoted uncertainties for these relative distance moduli arise from terms B \& C of our error budget, which is listed in detail in Table~\ref{tab:err}. For comparison, we also list the error budget typical of Cepheid distance determinations based on HST/WFPC2 observations \citep[e.g.,][]{gibson00} as well as the anticipated error budget after our follow-up NICMOS and ACS/HRC data are incorporated in the analysis and the uncertainty in the maser distance is reduced.

We derived relative distance moduli for the two fields using the methodology of \citet{freedman01}, in which one calculates mean $V$ and $I$ distance moduli for the Cepheid ensemble (i.e., neglecting differential reddening among Cepheids). We did not apply rejection criteria 2+3, since they were not used by those authors, and used the same period cuts as above. We obtained $\Delta\mu_V=\muvkpo, \Delta\mu_I=\muikpo$ and $\Delta\mu_0=\muokpo$~mag ($N=\nouk$, outer), and $\Delta\mu_V=\muvkpi,\Delta\mu_I=\muikpi$ and $\Delta\mu_0=\muokpi$~mag ($N=\nink$, inner). These values are consistent with a previous HST/WFPC2 Cepheid distance to \gal\ derived by \citet{newman01} using the same methodology. Their ALLFRAME photometry of $N=7$ Cepheids with $P=10-21$~d yielded $\Delta\mu_0=10.90\pm0.10_r\pm0.06_s$~mag.

\subsection{Metallicity dependence}
The two fields under study provide an excellent opportunity to obtain a differential measurement of the metallicity dependence of the Cepheid P-L relation. We adopted an abundance gradient for \gal\ measured by \citet{zaritsky94} and expressed in their ``empirical'' metallicity scale as:
\begin{equation}
[O/H] = 8.97\pm 0.06 - 0.49\pm 0.08 ( \rho - 0.4 )\ {\rm dex} \label{eqn:z}
\end{equation}
\noindent{where $\rho$ is the deprojected galactocentric radius, expressed as a fraction of the isophotal radius $\rho_0$:}
\begin{eqnarray}
x & = & (\alpha-\alpha_0) \cos{\phi} + (\delta - \delta_0) \sin{\phi} \\
y & = & \{(\delta - \delta_0) \cos{\phi} -(\alpha-\alpha_0) \sin{\phi}\} / (b/a) \\
\rho & = & (x^2+y^2)^{\frac{1}{2}} / \rho_0
\end{eqnarray}
We computed the deprojected galactocentric distances of the Cepheids using these equations. We adopted $\phi=149\arcdeg\!\!.75, b/a=0.413$ and $\rho_0=7\arcmin\!\!.76$ \citep[derived from a least-squares fit to the data in Table 2 of][]{zaritsky94} and a position for the center of \gal\ in J2000.0 coordinates of $(\alpha,\delta)=12^{\rm h}18^{\rm m}57\fracs5046$, $+47\arcdeg18\arcmin14\farcs303$ \citep{herrnstein05}.

Figure~\ref{fig:z} shows the correlation between true distance modulus and deprojected galactocentric distance, or its corresponding abundance according to Eqn.~\ref{eqn:z}. The sample plotted in this figure comprises all Cepheids from Table~\ref{tab:cephb} with $L_V>2$ (i.e., the restricted sample) and $P>6$~d (outer field) or $P>12$~d (inner field). At the suggestion of the referee, we further restricted the samples to ensure that they cover the same range of extinction, $0.05 \le \ebv \le 0.28$ mag (N=69 Cepheids).

A least-squares fit to the data yields {\boldmath $\gamma=\metz$}~{\bf mag dex \boldmath$^{-1}$} and {\boldmath $\Delta\mu_0$}{\bf(\gal\,-\,LMC)}{\boldmath $=\dmu$~{\bf mag}}, measured at $12+\log [O/H]=8.5$~dex. The best fit is represented by a solid line in Fig.~\ref{fig:z}. Since this is a differential measurement within a single galaxy, the random uncertainty arises from the scatter in the individual distance moduli and the systematic error is due to the uncertainty in the determination of the \citeauthor{zaritsky94} gradient.  Figure~\ref{fig:resz} shows the residuals of the individual distance moduli about the fit, plotted as a function of $\ebv$.

Our measurement compares favorably with the recent determination of \citet{sakai04}, who used the tip of the red giant branch as a fiducial distance indicator under the assumption that is unaffected by abundance differences. They derived $\gamma=-0.25\pm0.09$~mag dex$^{-1}$ by comparing distances determined using Cepheid variables and the Tip of the Red Giant Branch (hereafter TRGB) to 17 nearby galaxies. The Cepheid distances were calculated using the same P-L relations we adopted (Eqns.~\ref{eqn:plov} \& \ref{eqn:ploi}).

Our result is also consistent with, but more statistically significant than an earlier differential determination of the metallicity effect by \citet{kennicutt98}, who found $\gamma=-0.24\pm0.16$~mag dex$^{-1}$ based on {\it HST} observations of Cepheids in two fields within M101. 

Likewise, our findings are in agreement with the values of metallicity dependence derived by \citet{kochanek97} through an analysis of Cepheid magnitudes and colors in multiple galaxies, and by \citet{sasselov97} from a differential comparison of Large and Small Magellanic Cloud Cepheids. We find a difference in distance modulus between the inner and outer fields of $\delta\mu_0=-0.15\pm0.04$~mag for a mean abundance difference of $\Delta Z=0.45$~dex; the aforementioned studies would have predicted $\Delta\mu_0=-0.15\pm0.06$~mag and $-0.18\pm0.08$~mag, respectively.

Adopting the $T_e$ metallicity scale of \citet{kennicutt03}, the coefficient of the metallicity dependence becomes $\gamma=\mett$ mag dex$^{-1}$.

\subsection{A Tip of the Red Giant Branch distance to \gal}

At the suggestion of the referee, we determined a distance to \gal\ using the TRGB method \citep{lee93,sakai04}. The $I$-band master image of the outer field reaches a depth of $I\sim27$~mag, which is significantly deeper than the expected TRGB magnitude. The $V$-band master image reaches a depth of $V\sim28$~mag, which is sufficient to reject all stars in the I-band luminosity function with $\vi\le 1$~mag. Such a color cut is standard practice in TRGB studies \cite{sakai04,mendez02}.

The outer field $I$-band master object list (\S2.2.2, 2), contains $2.05\times 10^5$ objects. We rejected objects that appeared in less than half of the individual frames or exhibited signs of variability ($L_I>0.75$), reducing the sample to $1.37\times10^5$ objects. Then, we rejected a small fraction (2\%) of the remaining objects which exhibited a poor fit to a stellar PSF relative to other objects of the same magnitude. These are either faint galaxies or crowded stars. We matched the remaining $1.35\times 10^5$ objects against the $V$-band master list and rejected all objects with $\vi < 1$~mag. Thus, the final $I$-band luminosity function that served as input for the TRGB detection algorithm consisted of $1.2\times 10^5$ stars with $\vi>1$~mag.

We computed the TRGB magnitude following the procedures described in \citet{sakai96} and \citet{mendez02}. We computed a continuous luminosity function $\phi(m)$ using Equation (A1) of \citet{sakai96} and a logarithmic edge-detection function $E(m)$ using Equation (4) of \citet{mendez02}. We measured the TRGB magnitude  by identifying the highest peak in the product $E(m) \sqrt{\phi(m)}$ and fitting a cubic spline to the region $\pm0.15$~mag about the peak. Lastly, we estimated the uncertainty in our measurement of the TRGB magnitude by performing a bootstrap test with 500 simulations, as carried out by \citet{sakai04}.

The right panel of Figure~\ref{fig:trgb} shows the values of $\phi(m)$ and $E(m)$ that we obtained, resulting in a clear detection of the TRGB at $I_{TRGB}=25.42\pm0.02$~mag. For reference, the TRGB magnitude is also shown as a dashed line in the $I$-band CMD plotted in the left panel of Figure~\ref{fig:trgb}; note that the actual dataset used to measure the TRGB was far more complete than what can be shown in the CMD, containing $4\times$ more stars with $\vi>1$~mag and reaching $I\sim27$~mag.

We corrected the observed $I$ magnitude of the TRGB for foreground reddening (\S3.2) by $A_I=0.03$~mag. We also applied bolometric and metallicity corrections, following Equations (1)-(4) of \citet{sakai99}. These equations require the determination of the mean $\vi$ color of stars at the TRGB edge and 0.5~mag below it. We determined those values to be $(\vi)_{TRGB}=2\pm0.25$ and $(\vi)_{-3.5}=1.75\pm0.25$ by constructing histograms of the $\vi$ color distribution for stars within $\pm0.1$~mag of $I=25.42$ and $25.92$~mag, respectively. The bolometric and metallicity correction amounts to $+0.02\pm0.08$~mag

After these corrections, we find $I_{TRGB}^0=25.41\pm0.04_r\pm0.08_s$~mag. The corresponding value for the LMC \citep{sakai99} is $I_{TRGB}^0 (LMC)=14.54\pm0.04_r\pm0.06_s$~mag. Thus, we determine an LMC-relative distance modulus to \gal\, based on the TRGB method, of $\Delta\mu_{0,TRGB}=\dmutip$~mag, in excellent agreement with the Cepheid relative distance modulus obtained in \S4.3. Additionally, this determination allows us to increase the sample of galaxy fields used by \citet{sakai04} to determine the Cepheid metallicity dependence based on the observed difference between TRGB and Cepheid distance moduli. Figure~\ref{fig:z_trgb} is an updated version of the bottom panel of Figure 12 of \citet{sakai04}, with the addition of the two fields in \gal. The best-fit line to the data is $\gamma=-0.27\pm0.06$~mag dex$^{-1}$, in very good agreement with the metallicity dependence we independently derived in \S4.3.

\subsection{Other Period-Luminosity relations}

We considered in our analysis a second set of LMC Period-Luminosity relations derived by \citet{sandage04} using the sample of \citeauthor{udalski99} and additional long-period Cepheids ($P=10-80$~d) from the literature. These P-L relations have two slopes, with the break point set at $P=10$~d \citep[as motivated by][]{kanbur04}. We found no statistically significant difference between the distance moduli derived using the \citeauthor{udalski99} and the \citeauthor{sandage04} relations. This is consistent with the observation of \citet{ngeow05} that very large samples of Cepheids ($N> 10^2$) are required to detect the change in slope of the P-L relations.

Additionally, we considered the P-L relations derived by \citet{tammann03} for Milky Way Cepheids. In that study, the individual distance to each variable was derived using the Baade-Wesselink method and/or the open-cluster main-sequence fitting method. The authors determined P-L relations with slopes that were significantly steeper than those derived using LMC Cepheids. They attributed the change in slope to abundance differences, since the Milky Way Cepheids in their sample have a mean metallicity that is close to solar. Recently, \citet{saha06} recalibrated the peak luminosities of type Ia SNe using P-L relations whose slopes vary as a function of abundance and \citet{sandage06} used the results to derive $H_0=62\pm6$\ksm.

However, there is some controversy over the P-L relation slopes that are derived via the Baade-Wesselink method. \citet{gieren05} applied this technique to LMC Cepheids and derived different P-L slopes than those of \citet{udalski99}. They attributed the difference to a systematic error in the Baade-Wesselink technique, which requires the use of a period-dependent projection factor $p$. \citeauthor{gieren05} proposed a new $p$ factor that would resolve the discrepancy. However it still remains to be explained why \citeauthor{tammann03} derived essentially identical Milky Way P-L relations using a completely independent method (open cluster main-sequence fitting). Parallax measurements to Galactic Cepheids be provided by GAIA in the next decade may yield a definitive answer on this matter.

We can test the hypothesis of \citet{saha06} with our large sample of Cepheids in the inner field, since the application of the correct $V$ and $I$ P-L relations should lead to a distribution of distance moduli that is uncorrelated with period. We started with the restricted sample of 195 Cepheids in the inner field and excluded 23 objects with anomalous amplitude ratios (\S3.3.1), 24 variables with $P<6$~d, and 5 objects with $\mu_W$ outside $11\pm1$~mag. Next, we fit a slope to $\Delta\mu_0$ vs $P$ using an iterative least-absolute-deviation procedure with $3\sigma$ clipping, which rejected 6 outliers. Thus, our final sample consisted of 137 Cepheids. We carried out this exercise for three choices of P-L relation: \citet{udalski99}, \citet{sandage04}, and \citet{tammann03}. We tested the null hypothesis by computing the Spearman rank-order correlation coefficient $r_s$ for each choice of P-L relation. For comparison, we carried out the same exercise for the outer field sample and the P-L relations of \citeauthor{udalski99} Figure~\ref{fig:slp} shows the result of these tests.

The LMC P-L relations are a good fit to the samples of both fields. There is a small correlation for the inner field with $r_s=0.2\ (2.5\sigma)$, which decreases to $r_s=0.15\ (1.3\sigma$) if we use $P_{min}=12$~d as in \S4.2. The application of the Milky-Way P-L relations of \citet{tammann03} to the inner field sample yields a distribution that deviates noticeably from the null hypothesis, with $r_s=0.6\ (6.8\sigma)$. The correlation is still present, with $r_s=0.45\ (4\sigma)$, for $P_{min}=12$~d. Thus, we conclude that the LMC P-L relations are a better fit to both samples, regardless of their abundance difference.

\subsection{Implications for $H_0$ and $w$}
Since the mean abundance of LMC Cepheids ($12+\log[O/H]=8.5$~dex) lies within the range spanned by our sample of variables (Fig.~\ref{fig:z}) we have measured {\boldmath $\Delta\mu_0$}{\bf(\gal\,-\,LMC)}{\boldmath $=\dmu$}~{\bf mag} (\S4.3). Combined with the maser distance modulus to \gal, we infer the distance modulus of the LMC to be $\mu_0(LMC)=\mulmc$~mag. This corresponds to a distance of $D(LMC)=\dlmc$~kpc, which is in excellent agreement with the value of $48.3\pm1.4$~kpc derived from eclipsing binaries \citep[see Case II in Table 8 of][]{fitzpatrick03}. Importantly, both distance estimates are mainly geometric, independent of each other, and do not rely on any ``standard candles''.

We note that in the near future, there will be four galaxies with ``geometric distances'' that can serve as absolute calibrators for the Cepheid Distance Scale: the Large Magellanic Cloud \citep[with multiple DEB distances, see][and references therein]{fitzpatrick03}, Messier 31 \citep[with a DEB distance by][]{ribas05}, Messier 33 \citep[with a DEB distance by][]{bonanos06} and \gal\ \citep[with the maser distance by][]{humphreys07}. Thus, we can expect a significant reduction in the uncertainty of the ``first rung'' of the Extragalactic Distance Scale, which has been a dominant source of uncertainty in recent determinations of $H_0$.

The implied decrease in the distance to the LMC derived in this paper, relative to the adopted value of $D=50.1\pm2.3$~kpc \citep{freedman01,saha01a}, affects previously-derived values of $H_0$ by $\sim+3$\%. The increase in the coefficient of the metallicity dependence from $\gamma=-0.2\pm0.2$~mag dex$^{-1}$ \citep[adopted by][]{freedman01} to $\metz$~mag dex$^{-1}$ (\S 4.3) has an opposite effect on $H_0$ of $\sim-2\%$. As a result, the net effect on the calibration of secondary distance indicators is mitigated. Table 8 shows a re-calculation of the peak absolute $V$ magnitude of type Ia SNe recently determined by \citet{riess05}, which changes only by -0.03~mag to $M^0_V=-19.14\pm0.06$~mag. The resulting value of $H_0$ is $\hnod$\ksm.

Recently, \citet{spergel06} presented a determination of cosmological parameters based on 3 years of WMAP observations. CMB observations cannot provide strong constraints on the value of $H_0$ on their own, due to degeneracies in parameter space \citep{tegmark04}. Figure~\ref{fig:wmap} shows the degeneracy in the $\Omega_M-w$ plane. The addition of an independent of $H_0$ from Cepheids significantly reduces that degeneracy \citep{hu05}.

We calculated the improvement due to a prior on $H_0$ (solid contours of Fig.~\ref{fig:wmap}) by resampling the Monte Carlo Markov Chains kindly made available by the WMAP team, using Eq.~B4 of \citet{lewis02}. We also calculated marginalized probability distributions for $w$ for increasingly more accurate priors on $H_0$. The results, which are shown in Figure~\ref{fig:h0w}, indicate that a 5\% prior on $H_0$ would reduce the $1\sigma$ uncertainty in $w$ to $\pm0.1$. As shown by \citeauthor{spergel06}, the combination of CMB data with more than one prior (e.g., Cepheids, type Ia SNe and large-scale structure) can further refine the measurement of $w$.

A determination of $H_0$ to 5\% (see Table~\ref{tab:err}) is a conservative goal for the near term. It will require the re-estimation of a maser distance to \gal\ \citep{humphreys07}, the analysis of follow-up observations of the Cepheids discovered in this paper with other {\it HST} instruments \citep{bersier07,macri07a}, and the inclusion in the Cepheid sample of longer-period (40~d $<P<$90~d) variables discovered with GMOS on Gemini North \citep{macri07b}.

Further improvement on the accuracy of $H_0$, down to 1\%, may be obtained through maser distances to a large number of galaxies in the Hubble flow, which could be discovered with the Square Kilometer Array and its prototypes \citep{greenhill04}.

\section{Conclusions}

The five main results presented in this paper are the following:

\vspace*{3pt}

\noindent {\bf 1.}\ We discovered $\ntot$ Cepheid variables in two fields located within the galaxy \gal, with accurately calibrated $\bvi$ photometry in twelve epochs per band.

\noindent {\bf 2.}\ We determined a relative distance modulus between \gal\ and the Large Magellanic Cloud, based on Cepheid variables, of $\Delta\mu_0=\dmu$~mag.

\noindent {\bf 3.}\ We determined a relative distance modulus between these two galaxies, based on the Tip of the Red Giant Branch method, of $\Delta\mu_{0,TRGB}=10.87\pm0.04_r$~mag.

\noindent {\bf 4.}\ We measured a metallicity dependence of the Cepheid distance scale of $\gamma=\metz$~mag dex$^{-1}$.

\noindent {\bf 5.}\ Our observations are best fit with P-L relations that do not invoke changes in slope as a function of abundance.

\acknowledgments{We thank the Telescope Allocation Committee of the {\em Hubble Space Telescope} for granting telescope time for this project in Cycle 12. We were partially supported by HST Grant HST-GO-09810, provided by NASA through a grant from the Space Telescope Science Institute, which is operated by the Association of Universities for Research in Astronomy, Incorporated, under NASA contract NAS5-26555. Support for L.M.M. was provided by NASA through Hubble Fellowship grant HST-HF-01153 from the Space Telescope Science Institute and by the National Science Foundation through a Goldberg Fellowship from the National Optical Astronomy Observatory.

Some of the data presented in this paper were obtained from the Multimission Archive at the Space Telescope Science Institute (MAST). We acknowledge the use of the Canadian Astronomy Data Centre, which is operated by the Herzberg Institute of Astrophysics, National Research Council of Canada. This research has made use of the NASA/IPAC Extragalactic Database (NED) which is operated by the Jet Propulsion Laboratory, California Institute of Technology, under contract with the National Aeronautics and Space Administration. We acknowledge the use of the Legacy Archive for Microwave Background Data Analysis (LAMBDA), which is supported by the NASA Office of Space Science. We thank the WMAP team for making their data publicly available. This research has made use of NASA's Astrophysics Data System Bibliographic Services.

L.M.M. wishes to thank Arjun Dey, Shashi Kanbur, Chris Kochanek, Tom Matheson, Jeremy Mould, Adam Riess, Abi Saha, Peter Stetson and Mat\'\i as Zaldarriaga for helpful discussions and comments during the preparation of this paper. We thank the referee, Shoko Sakai, for her helpful comments and suggestions during the review of this manuscript.}

\ \par

Due to size limitations imposed by arxiv.org, print-quality Figures appear only in the full-resolution PDF and PS versions, which are available at {\tt http://www.noao.edu/staff/lmacri/0608211-full.pdf} or {\tt http://www.noao.edu/staff/lmacri/0608211-full.ps.gz}, respectively.

\LongTables

\clearpage

\ \par\vfill

\vfill\clearpage\ \par\vfill

\vfill\clearpage\ \par\vfill

\vfill\clearpage\ \par\vfill

\vfill

\vfill\clearpage\ \par\vfill

\vfill\clearpage\ \par\vfill

\vfill\clearpage\ \par\vfill

\newcounter{subfig}
\setcounter{subfig}{1}
\renewcommand{\thefigure}{\arabic{figure}\alph{subfig}}
\begin{figure}[ht]
\center{\includegraphics[width=6.5in]{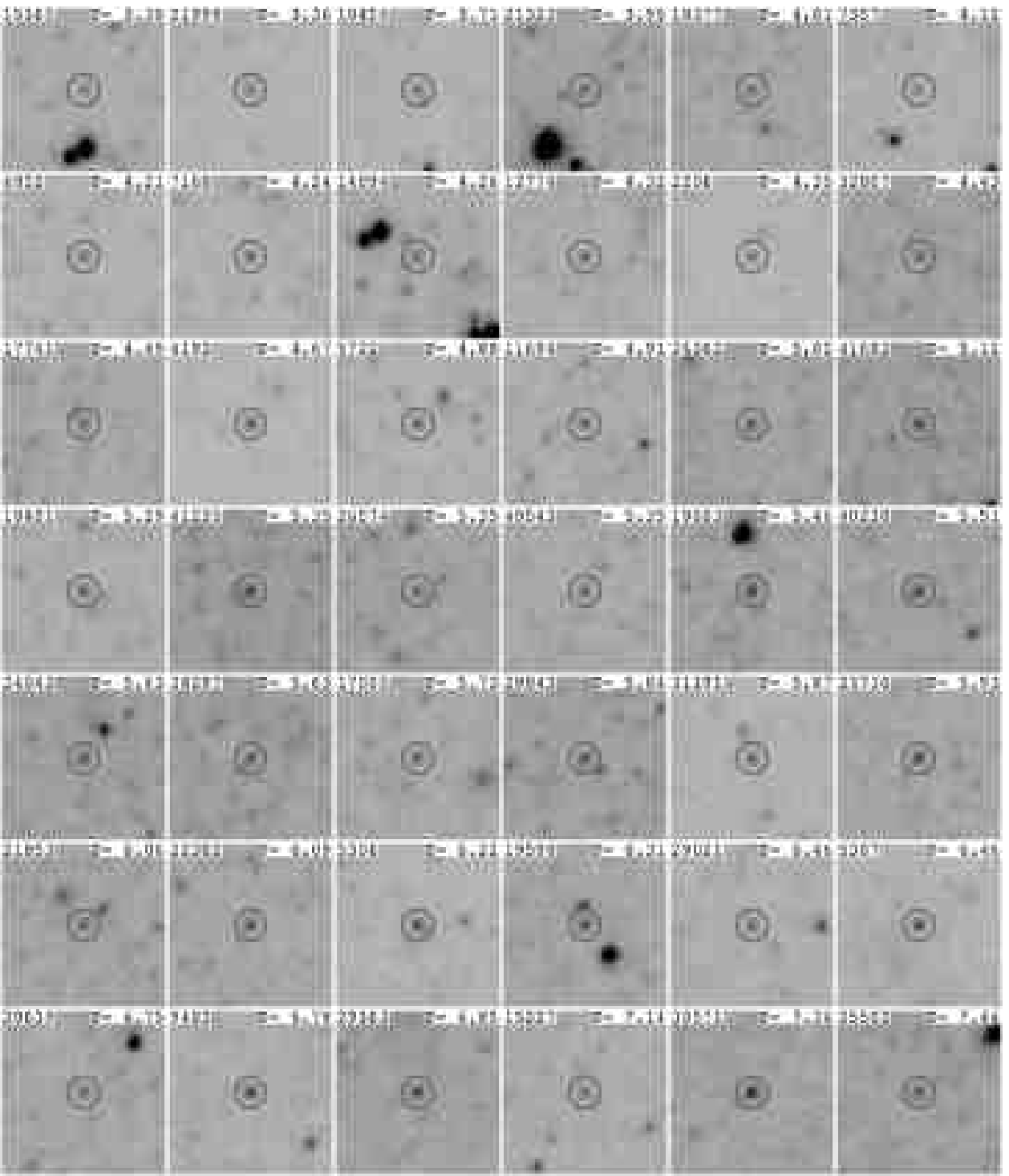}}
\caption{Individual finding charts for the Cepheids listed in Table~\ref{tab:cephb}. Each box is $2.5\arcsec$ on a side.}
\label{fig:fchart}
\end{figure}

\vfill\clearpage\ \par\vfill

\addtocounter{figure}{-1}
\addtocounter{subfig}{1}
\begin{figure}[ht]
\center{\includegraphics[width=6.5in]{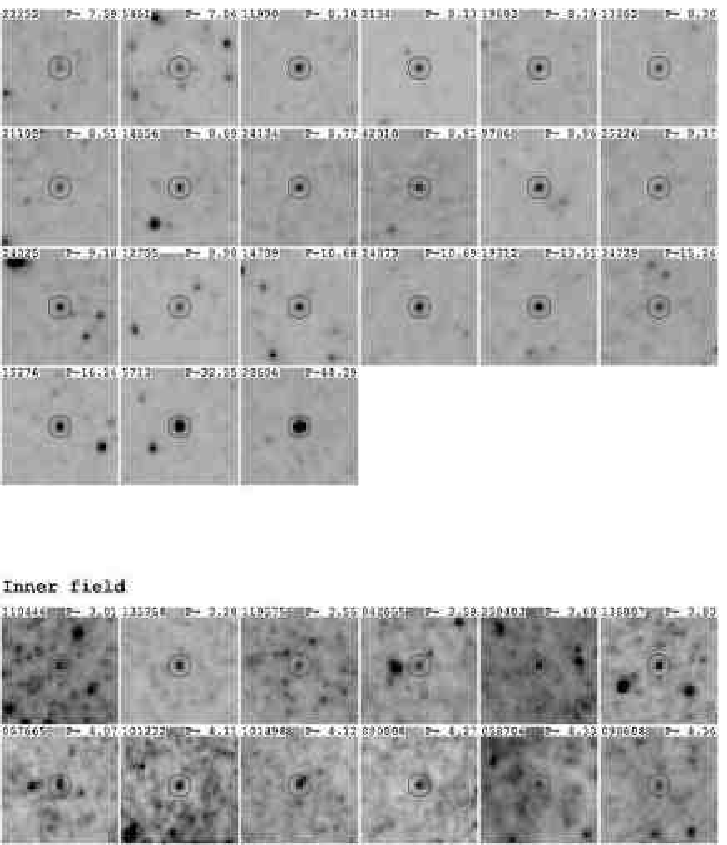}}
\caption{Individual finding charts for the Cepheids listed in Table~\ref{tab:cephb}. Each box is $2.5\arcsec$ on a side.}
\end{figure}

\vfill\clearpage\ \par\vfill

\addtocounter{figure}{-1}
\addtocounter{subfig}{1}
\begin{figure}[ht]
\center{\includegraphics[width=6.5in]{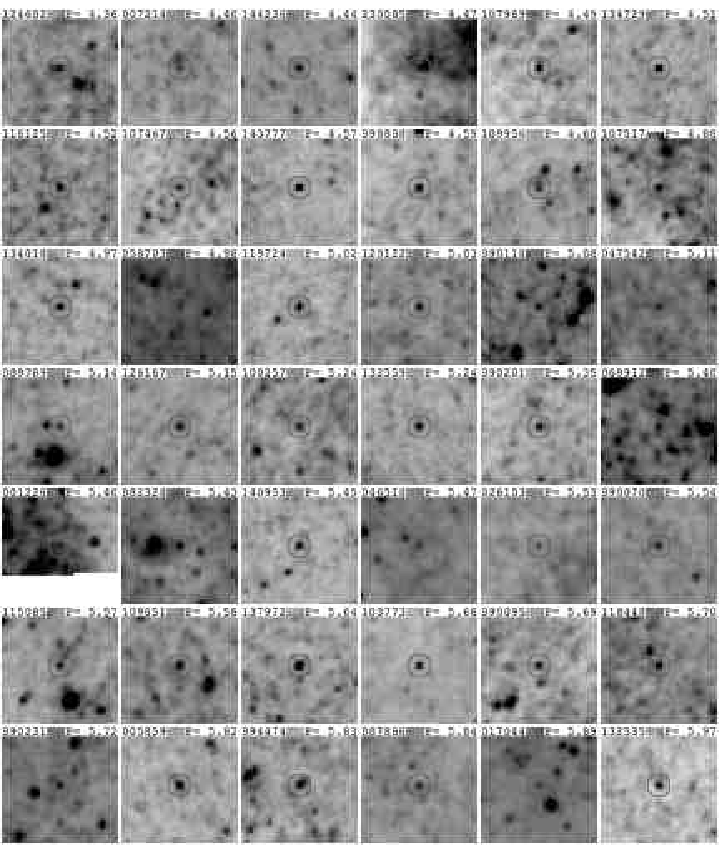}}
\caption{Individual finding charts for the Cepheids listed in Table~\ref{tab:cephb}. Each box is $2.5\arcsec$ on a side.}
\end{figure}

\vfill\clearpage\ \par\vfill

\addtocounter{figure}{-1}
\addtocounter{subfig}{1}
\begin{figure}[ht]
\center{\includegraphics[width=6.5in]{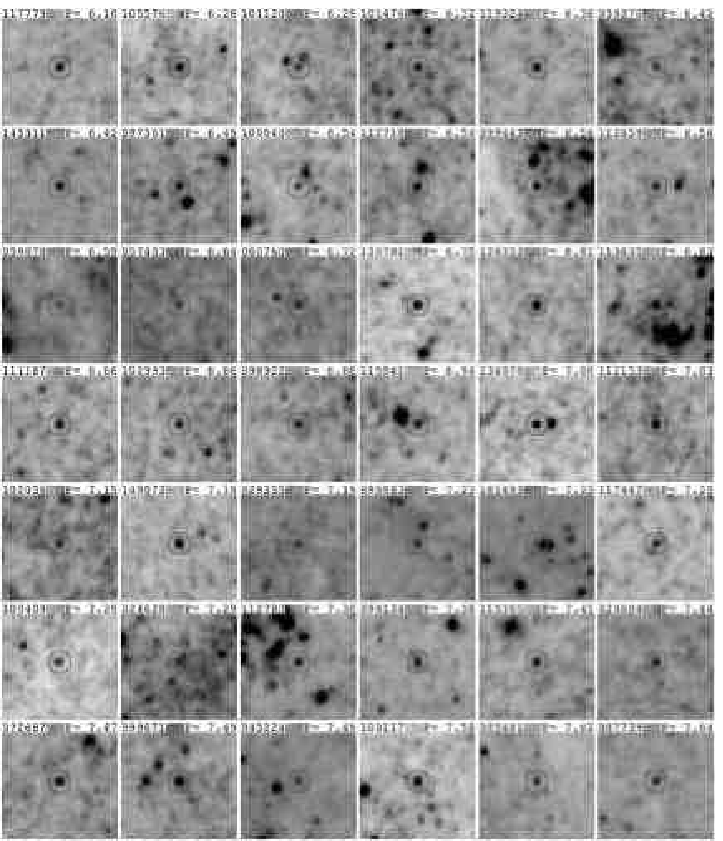}}
\caption{Individual finding charts for the Cepheids listed in Table~\ref{tab:cephb}. Each box is $2.5\arcsec$ on a side.}
\end{figure}

\vfill\clearpage\ \par\vfill

\addtocounter{figure}{-1}
\addtocounter{subfig}{1}
\begin{figure}[ht]
\center{\includegraphics[width=6.5in]{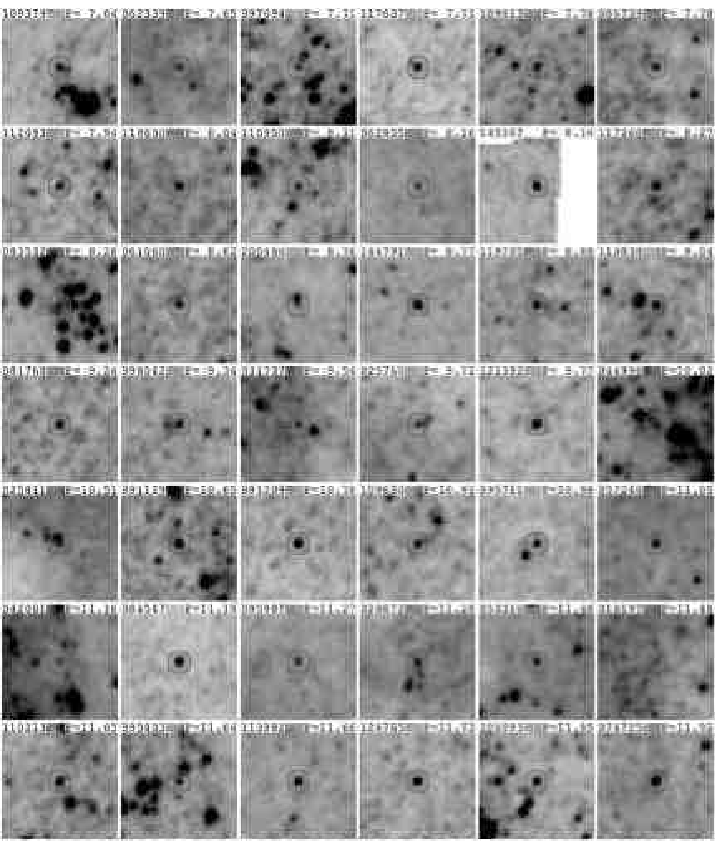}}
\caption{Individual finding charts for the Cepheids listed in Table~\ref{tab:cephb}. Each box is $2.5\arcsec$ on a side.}
\end{figure}

\vfill\clearpage\ \par\vfill

\addtocounter{figure}{-1}
\addtocounter{subfig}{1}
\begin{figure}[ht]
\center{\includegraphics[width=6.5in]{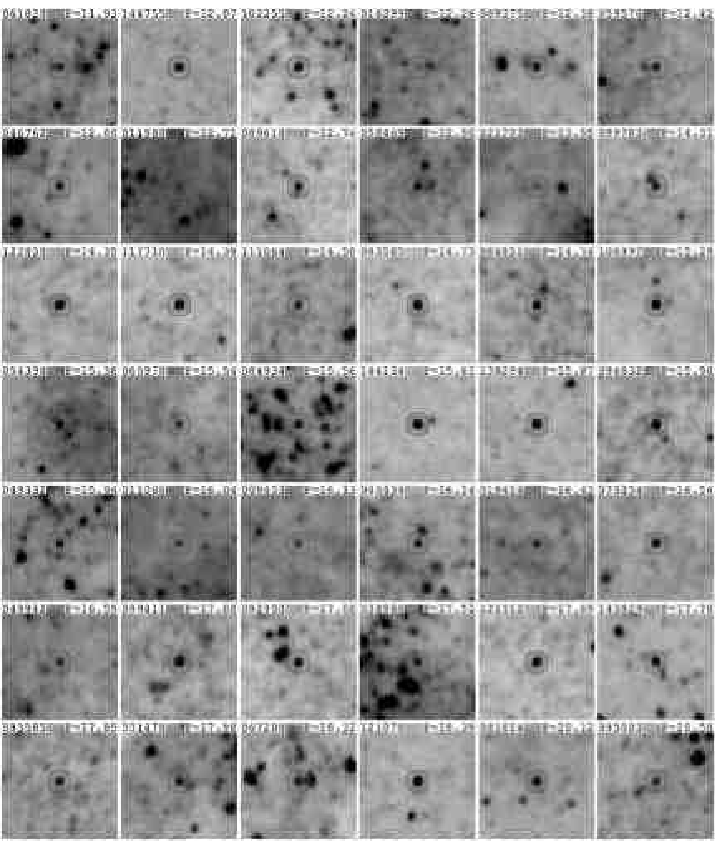}}
\caption{Individual finding charts for the Cepheids listed in Table~\ref{tab:cephb}. Each box is $2.5\arcsec$ on a side.}
\end{figure}

\vfill\clearpage\ \par\vfill

\addtocounter{figure}{-1}
\addtocounter{subfig}{1}
\begin{figure}[ht]
\center{\includegraphics[width=6.5in]{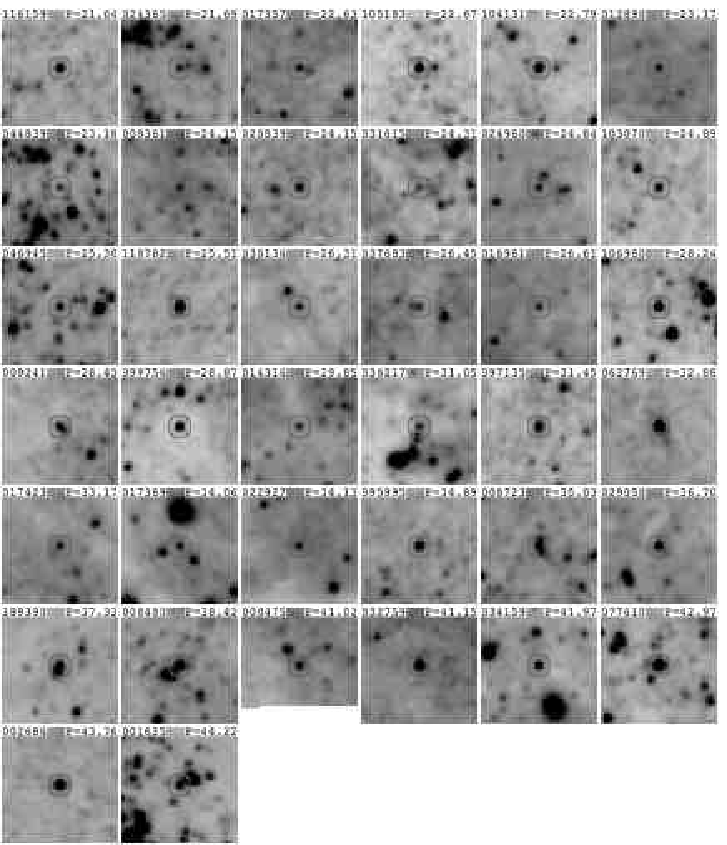}}
\caption{Individual finding charts for the Cepheids listed in Table~\ref{tab:cephb}. Each box is $2.5\arcsec$ on a side.}
\end{figure}

\vfill\clearpage\ \par\vfill

\renewcommand{\thefigure}{\arabic{figure}}
\begin{figure}[ht]
\center{\includegraphics[width=6.5in]{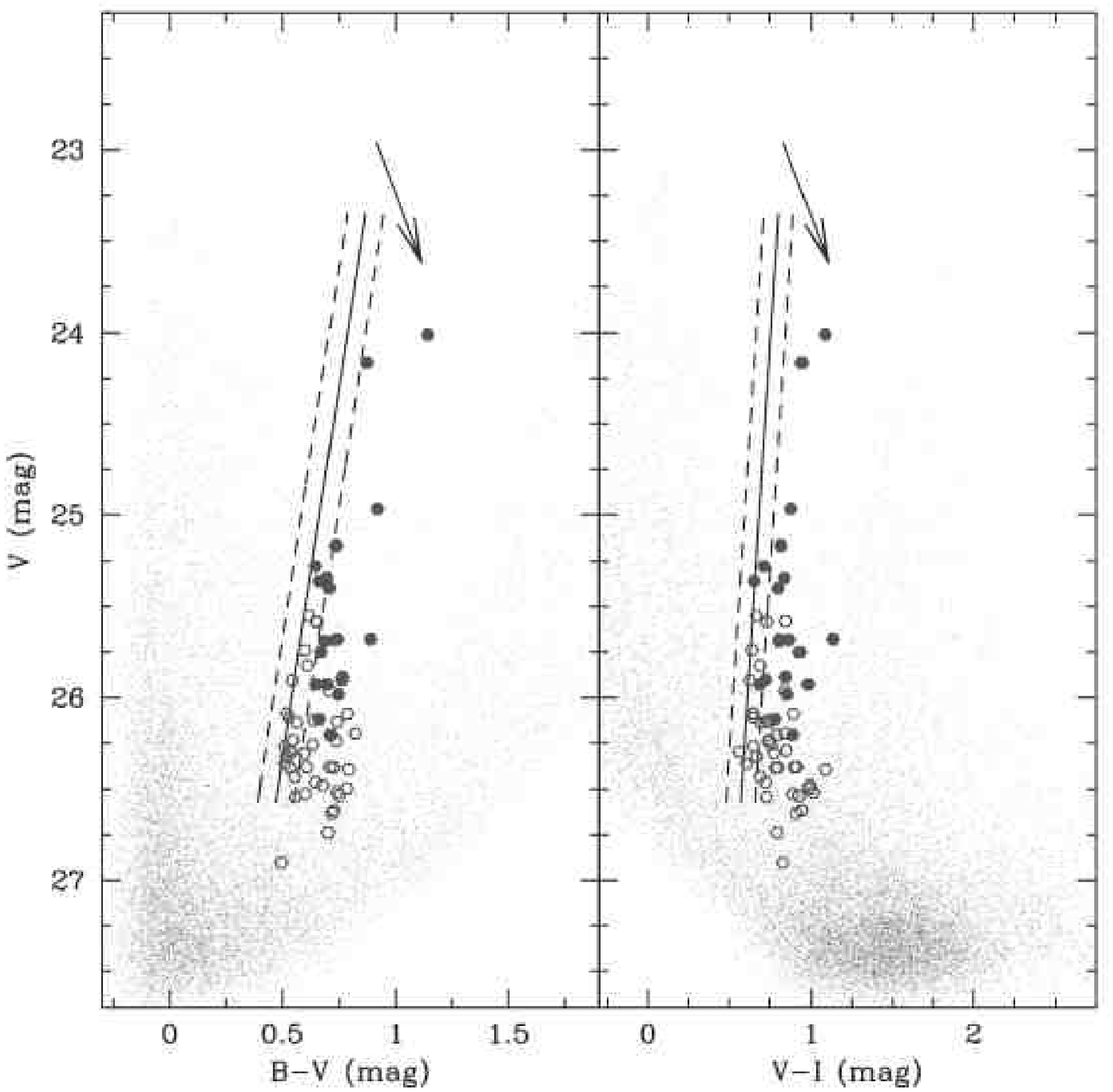}}
\caption{Color-magnitude diagrams for objects in the outer field of \gal. Cepheids are plotted using filled and open symbols for the restricted and extended samples, respectively. Field stars are represented by small dots. The dashed lines represent the zero-reddening instability strip of LMC Cepheids and its $2\sigma$ width. The arrows indicate the effect of $\ebv=0.2$~mag.}
\label{fig:cmdou}
\end{figure}

\vfill\clearpage\ \par\vfill

\begin{figure}[ht]
\center{\includegraphics[width=6.5in]{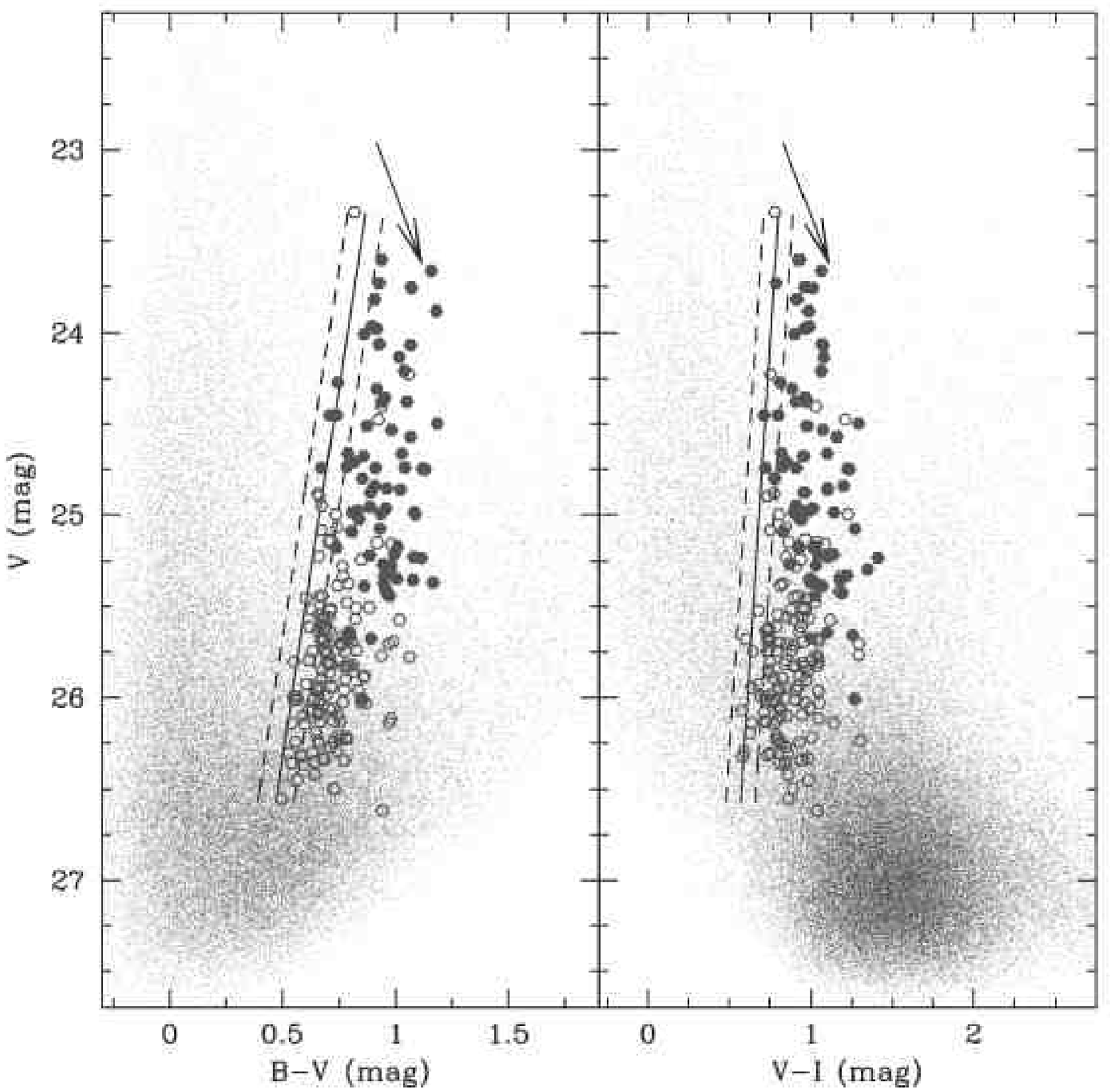}}
\caption{Color-magnitude diagrams for objects in the inner field of \gal. Cepheids are plotted using filled and open circles for the restricted and extended samples, respectively. Field stars are represented by small dots. The dashed lines represent the zero-reddening instability strip of LMC Cepheids and its $2\sigma$ width. The arrows indicate the effect of $\ebv=0.2$~mag.}
\label{fig:cmdin}
\end{figure}

\vfill\clearpage\ \par\vfill

\begin{figure}[ht]
\center{\includegraphics[width=6.5in]{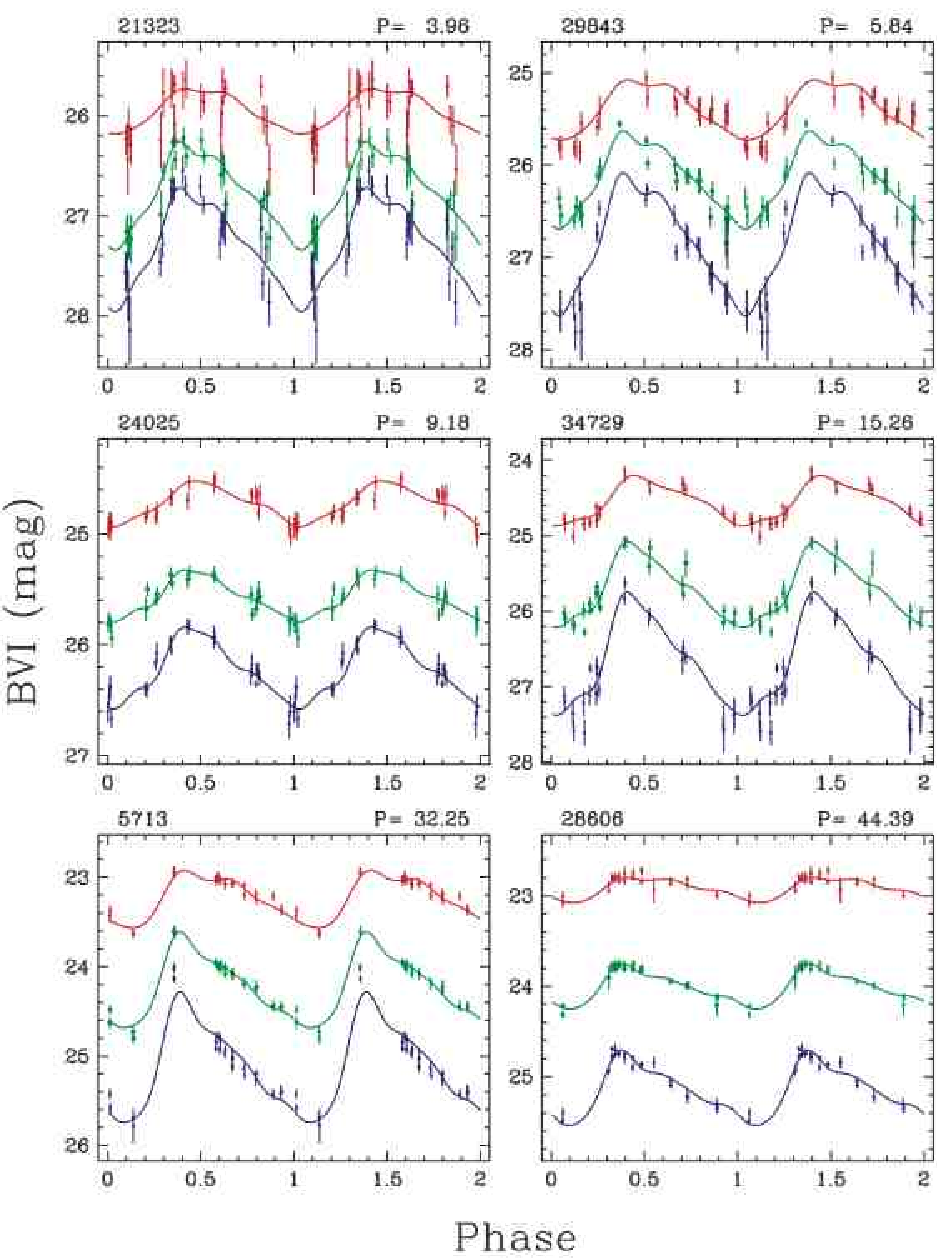}}
\caption{Representative light curves of Cepheids in the outer field. Blue: $B$; green: $V$; red: $I$. The solid lines indicate the best-fit light curve template from \citet{stetson96}.}
\label{fig:lcou}
\end{figure}

\vfill\clearpage\ \par\vfill

\begin{figure}[ht]
\center{\includegraphics[width=6.5in]{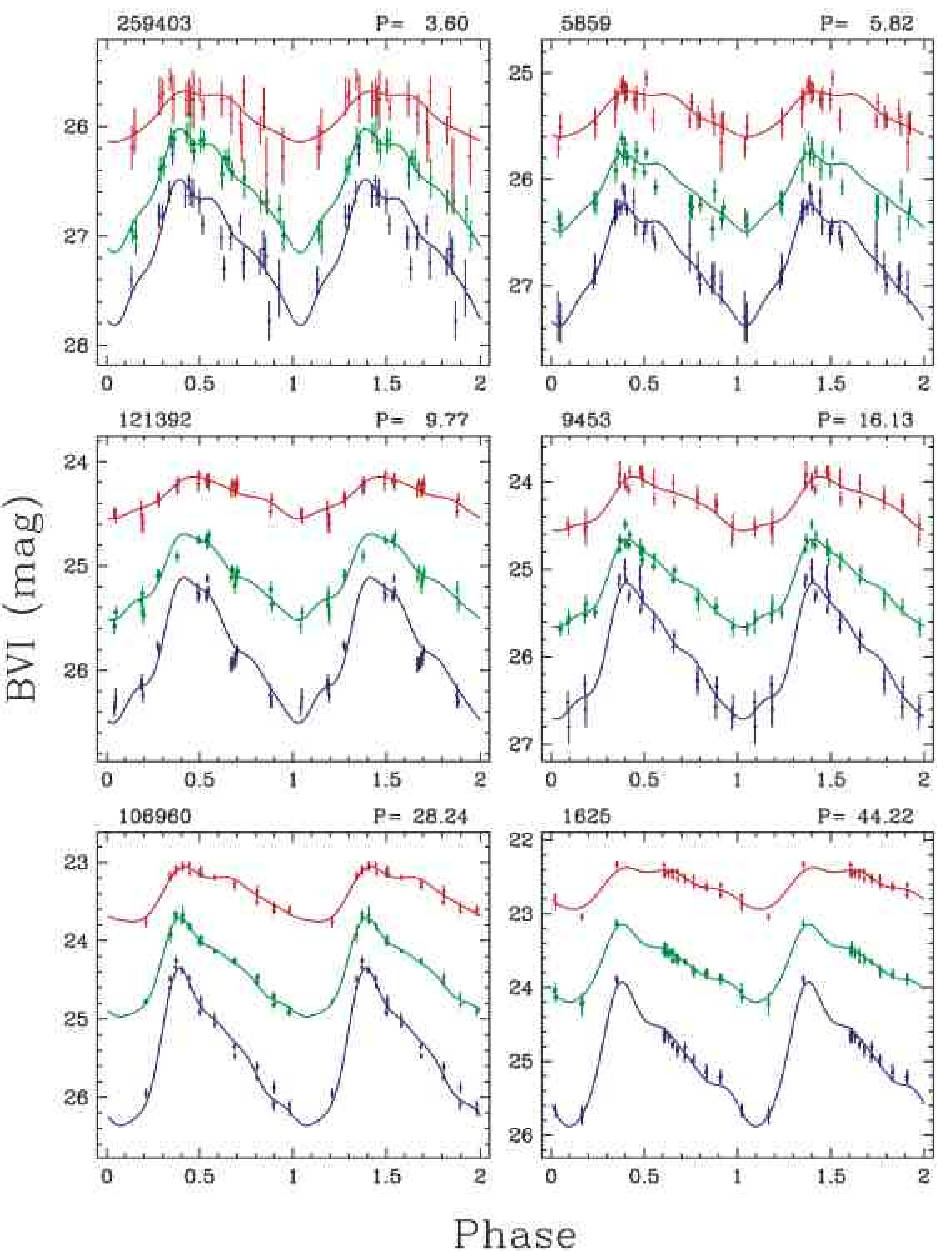}}
\caption{Representative light curves of Cepheids in the inner field. Blue: $B$; green: $V$; red: $I$. The solid lines indicate the best-fit light curve template from \citet{stetson96}.}
\label{fig:lcin}
\end{figure}

\vfill\clearpage\ \par\vfill

\begin{figure}[ht]
\center{\includegraphics[width=6.5in]{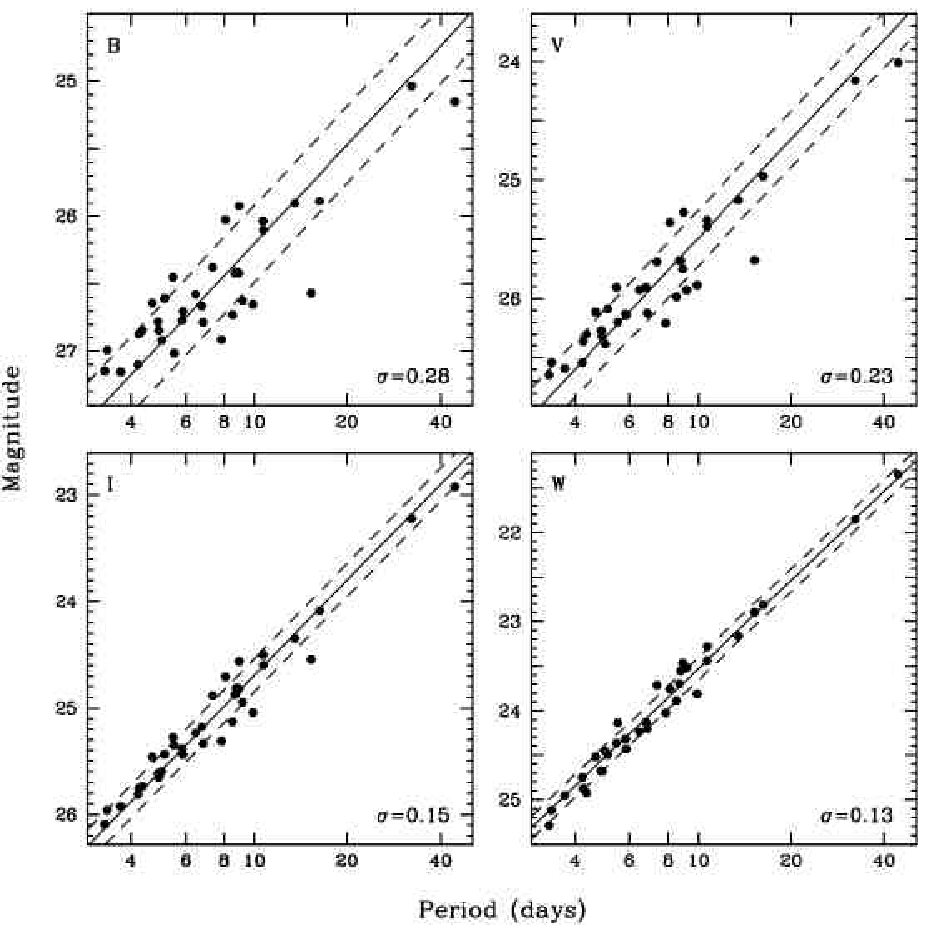}}
\caption{$\bvi$ and Wesenheit Period-Luminosity relations for the $L_V>2$ sample of Cepheids in the outer field. The solid lines represent the LMC P-L relations derived by \citet{udalski99}, shifted to the appropriate mean relative distance modulus. The dashed lines indicate the $1\sigma$ dispersion of the sample.}
\label{fig:plou}
\end{figure}

\vfill\clearpage\ \par\vfill

\begin{figure}[ht]
\center{\includegraphics[width=6.5in]{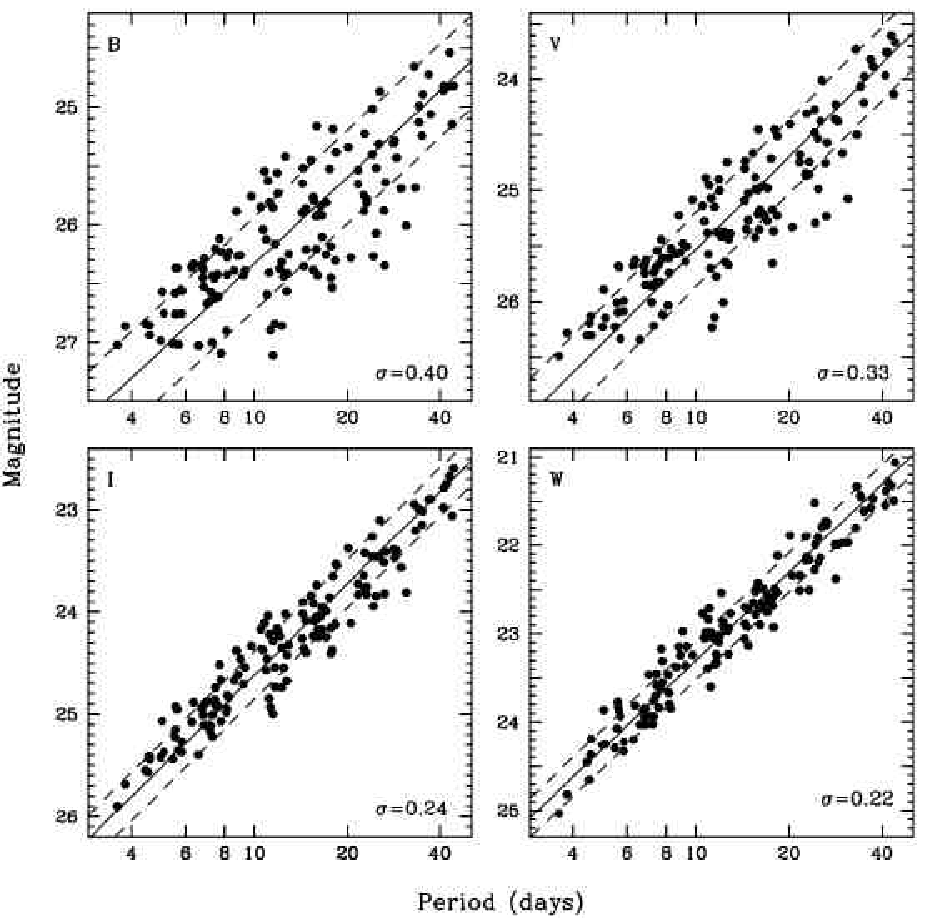}}
\caption{$\bvi$ and Wesenheit Period-Luminosity relations for the $L_V>2$ sample of Cepheids in the inner field. The solid lines represent the LMC P-L relations derived by \citet{udalski99}, shifted to the appropriate mean relative distance modulus. The dashed lines indicate the $2\sigma$ dispersion.}
\label{fig:plin}
\end{figure}

\vfill\clearpage\ \par\vfill

\begin{figure}[ht]
\center{\includegraphics[width=6.5in]{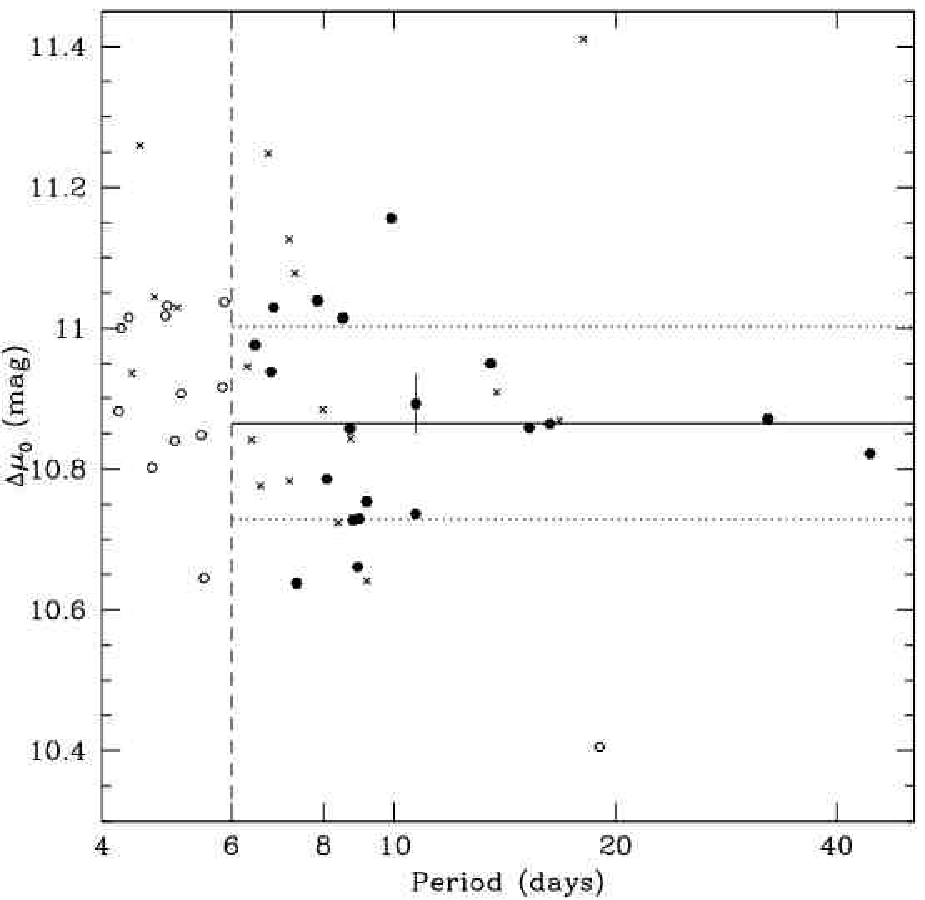}}
\caption{Relative distance modulus versus period for the restricted sample of Cepheids in the outer field. Crosses: Cepheid candidates that were rejected by selection criteria 1-4. Open circles: Candidates rejected by period cut or distance modulus clipping. Filled circles: Final sample of Cepheids used to determine the mean relative distance modulus. Solid line: mean relative distance modulus. Dotted lines: $1\sigma$ dispersion of the final sample. Dashed line: Final adopted period cut. A typical error bar is shown on one of the data points.}
\label{fig:dmou}
\end{figure}

\vfill\clearpage\ \par\vfill

\begin{figure}[ht]
\center{\includegraphics[width=6.5in]{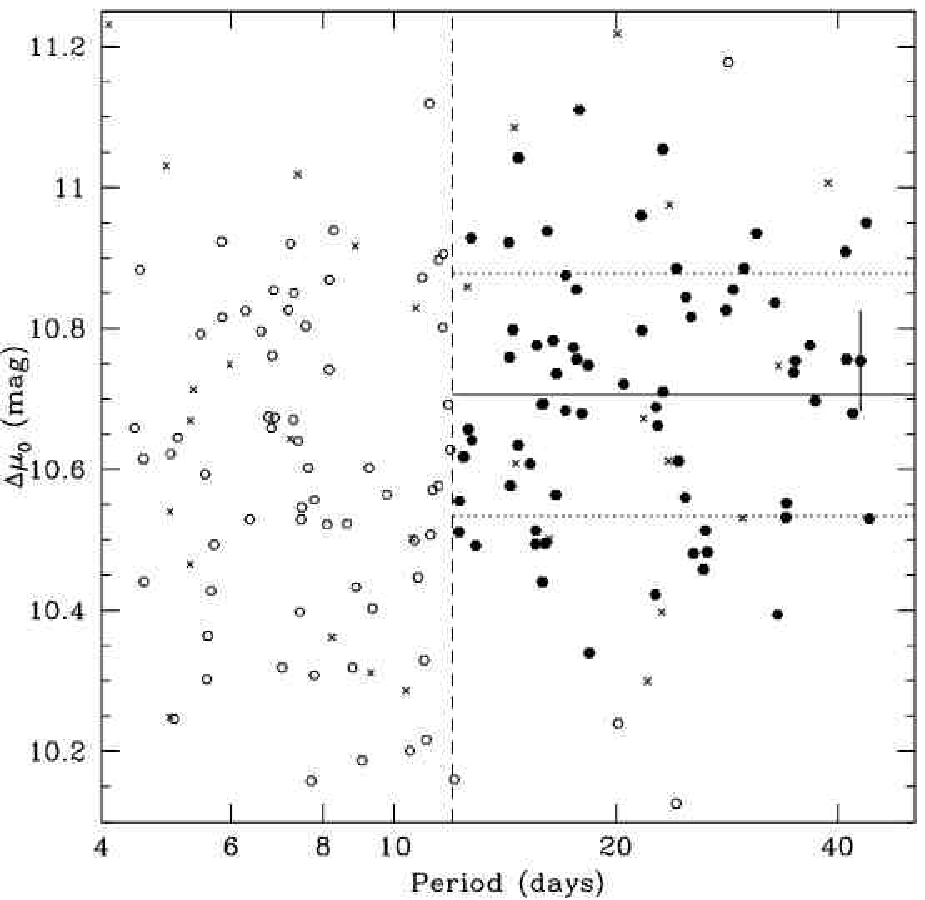}}
\caption{Relative distance modulus versus period for the restricted sample of Cepheids in the inner field. Crosses: Cepheid candidates that were rejected by selection criteria 1-4. Open circles: Candidates rejected by period cut or distance modulus clipping. Filled circles: Final sample of Cepheids used to determine the mean relative distance modulus. Solid line: mean relative distance modulus. Dotted lines: $1\sigma$ dispersion of the final sample. Dashed line: Final adopted period cut. A typical error bar is shown on one of the data points.}
\label{fig:dmin}
\end{figure}

\vfill\clearpage\ \par\vfill

\begin{figure}[ht]
\center{\includegraphics[width=6.5in]{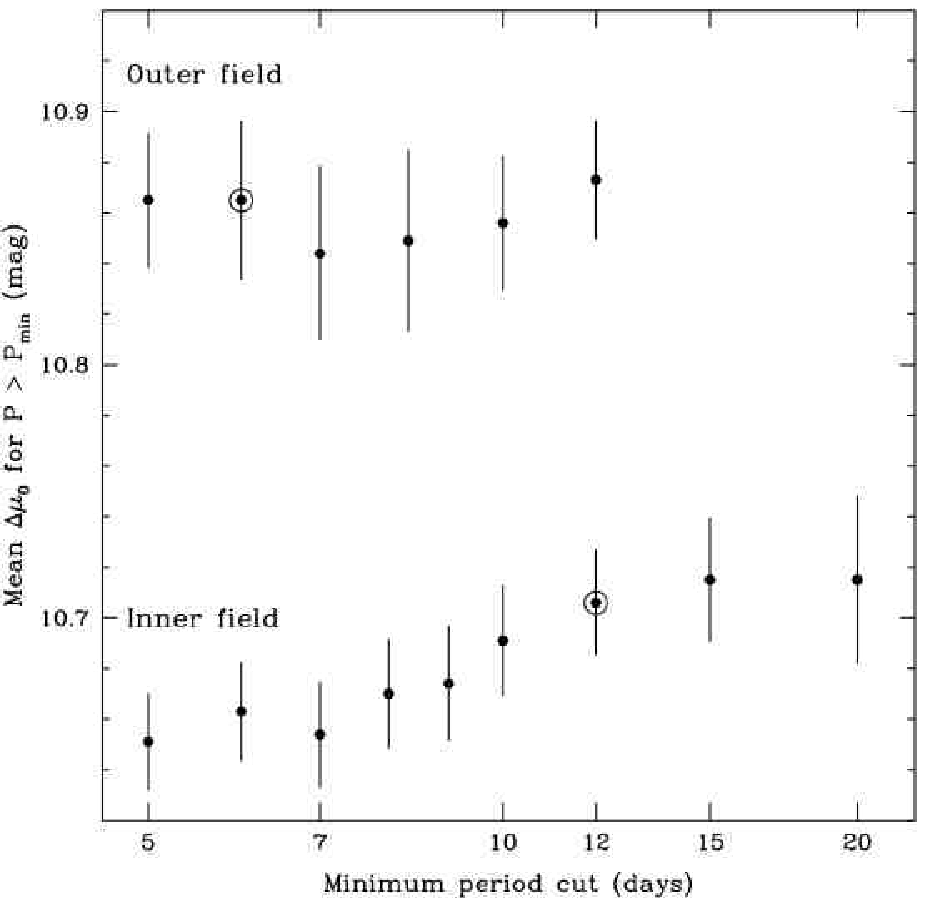}}
\caption{Mean distance modulus (relative to the LMC) as a function of cut at minimum period, for the $L_V>2$ samples of the outer (top) and inner (bottom) fields. The error bars represent the $1\sigma$ uncertainty in the mean. Our final choices for minimum period cut are indicated with open circles.}
\label{fig:dmp}
\end{figure}

\vfill\clearpage\ \par\vfill

\begin{figure}[ht]
\center{\includegraphics[angle=-90,scale=1.2]{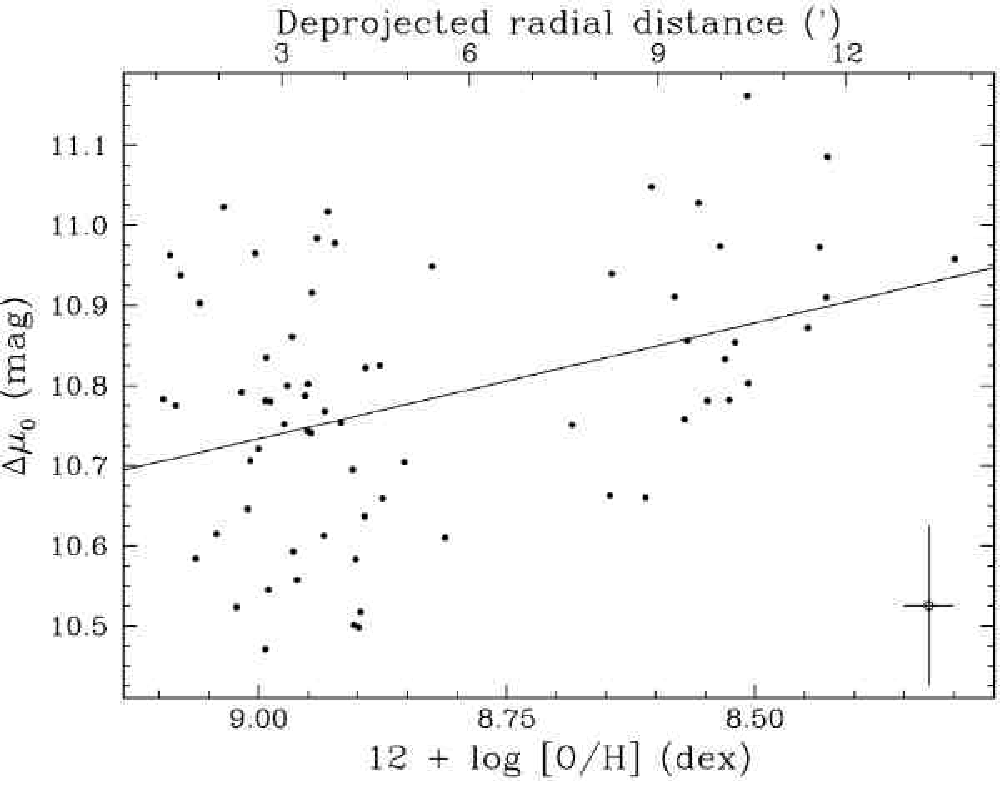}}
\caption{Correlation between distance moduli of individual Cepheids and their abundances, given by their galactocentric distances and the abundance gradient of \citet{zaritsky94}. The best-fit line has a value of $\metz$~mag dex$^{-1}$. A representative individual uncertainty is shown on the open symbol in the bottom right.}
\label{fig:z}
\end{figure}

\vfill

\begin{figure}[ht]
\center{\includegraphics[angle=-90,scale=1.2]{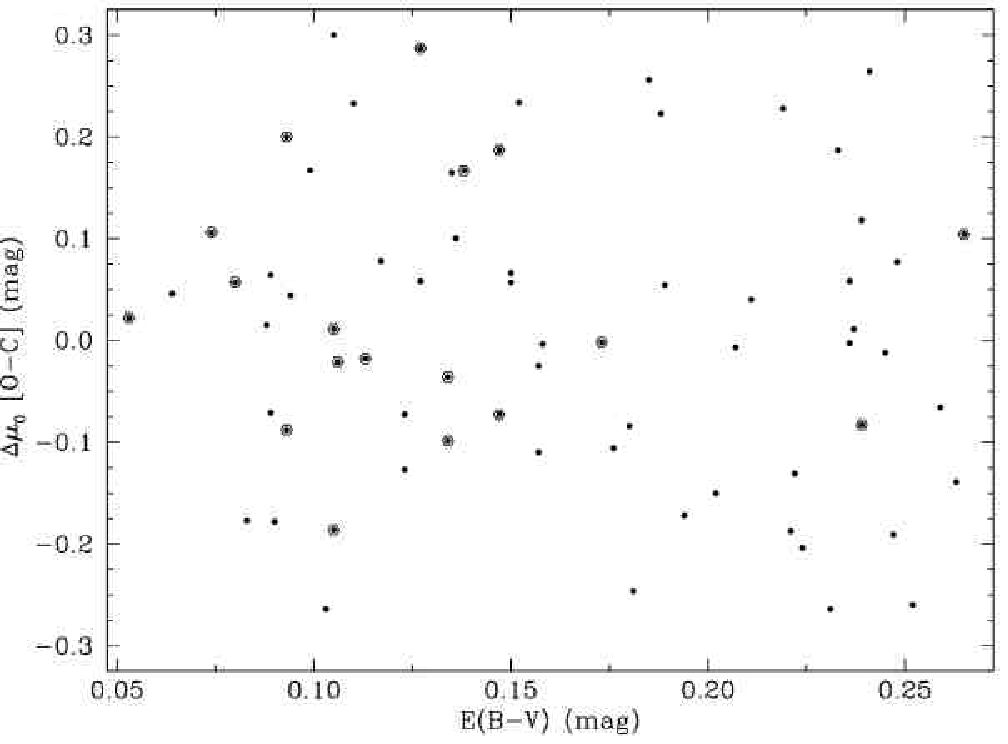}}
\caption{Residual of the individual Cepheid distance moduli about the best-fit line of Figure~\ref{fig:z}, plotted as a function of $\ebv$. Cepheids located in the outer field are represented by concentric open and filled symbols. Cepheids located in the inner field are indicated by filled symbols.}
\label{fig:resz}
\end{figure}

\vfill\clearpage\ \par\vfill

\begin{figure}[ht]
\center{\includegraphics[angle=-90,scale=1.2]{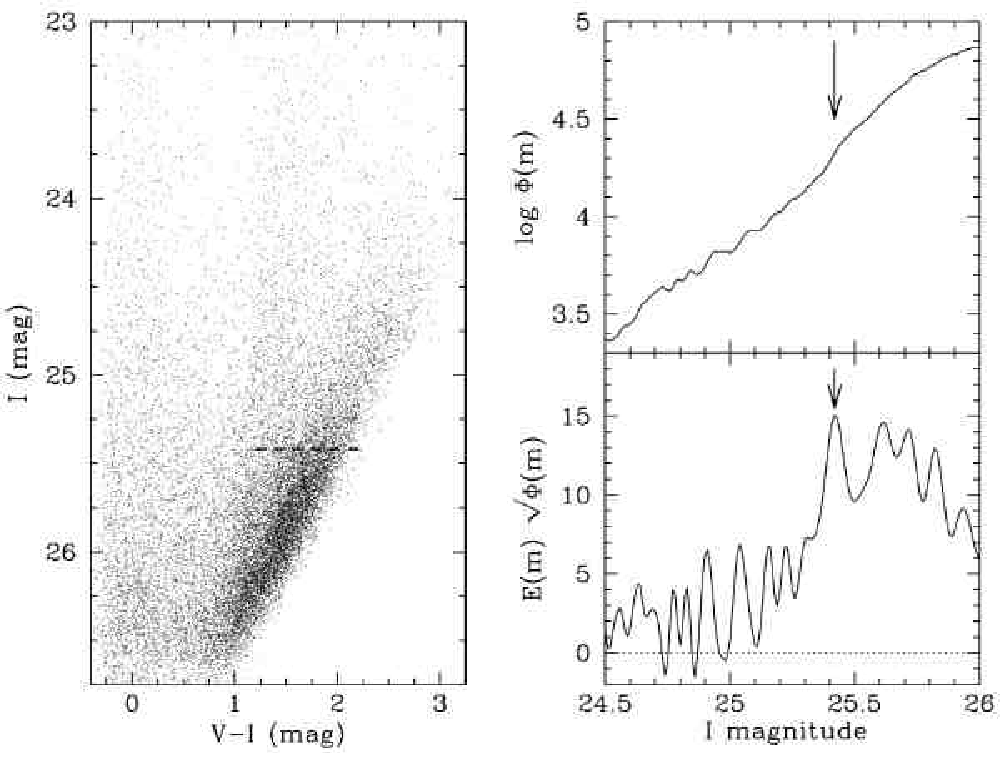}}
\caption{Determination of the $I$-band magnitude of the Tip of the Red Giant Branch in the outer field of \gal. Left: $I$-band color-magnitude diagram, indicating the location of the TRGB (dashed line). Right: $I$-band luminosity function $\phi(m)$ and edge function $E(m)\sqrt{\phi(m)}$, indicating the detection of the TRGB edge at $I=25.42\pm0.02$~mag.}
\label{fig:trgb}
\end{figure}

\vfill

\begin{figure}[ht]
\center{\includegraphics[angle=-90,scale=1.2]{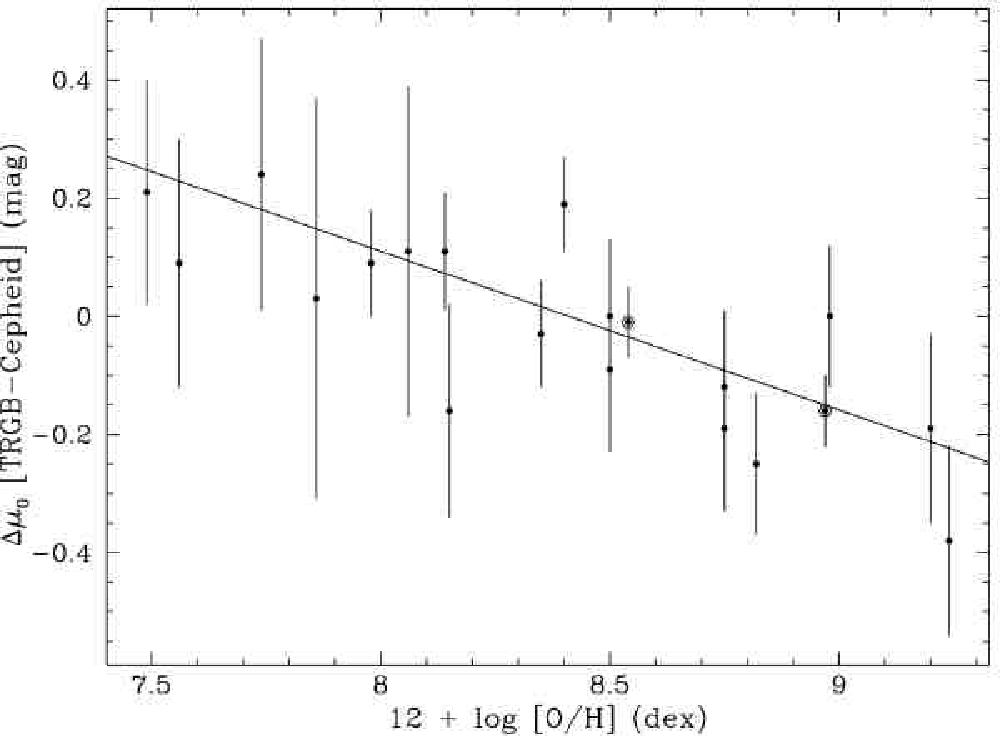}}
\caption{Cepheid metallicity dependence determined through a comparison of TRGB and Cepheid distance moduli for 20 fields in 18 galaxies. This Figure reproduces the bottom panel of Figure 12 of \citet{sakai04}, with the addition of two points for our fields in \gal, shown with concentric open and filled symbols.}
\label{fig:z_trgb}
\end{figure}

\vfill\clearpage\ \par\vfill

\begin{figure}[ht]
\center{\includegraphics[height=7in]{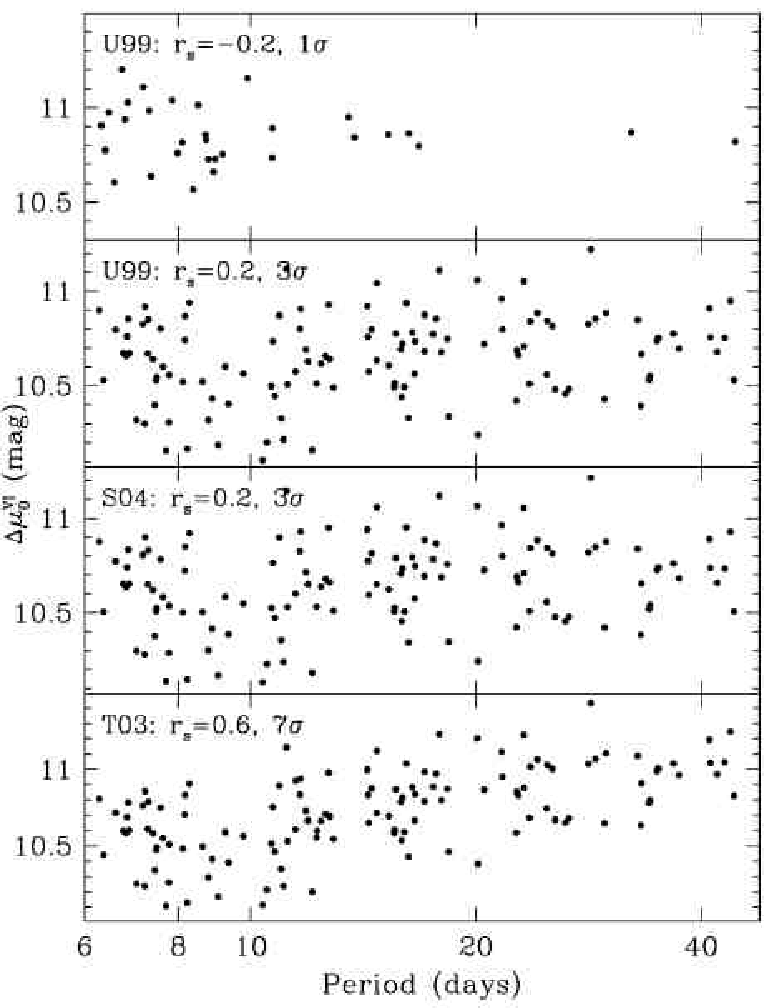}}
\caption{Correlation between period and extinction-corrected LMC-relative distance moduli for different choices of input P-L relation. Top panel: Outer field Cepheids. Bottom three panels: Inner field Cepheids. U99=\citet{udalski99}; S04=\citet{sandage04}; T03=\citet{tammann03}. The LMC P-L relations are good fits to the Cepheid samples of both fields. The adoption of the Milky-Way P-L relations of \citet{tammann03} leads to a residual slope with a significance of $\sim 7\sigma$~level for $P_{min}=6$~d or $\sim 4\sigma$ for $P_{min}=12$~d.}
\label{fig:slp}
\end{figure}

\vfill\clearpage\ \par\vfill

\begin{figure}[ht]
\center{\includegraphics[angle=-90,scale=1.2]{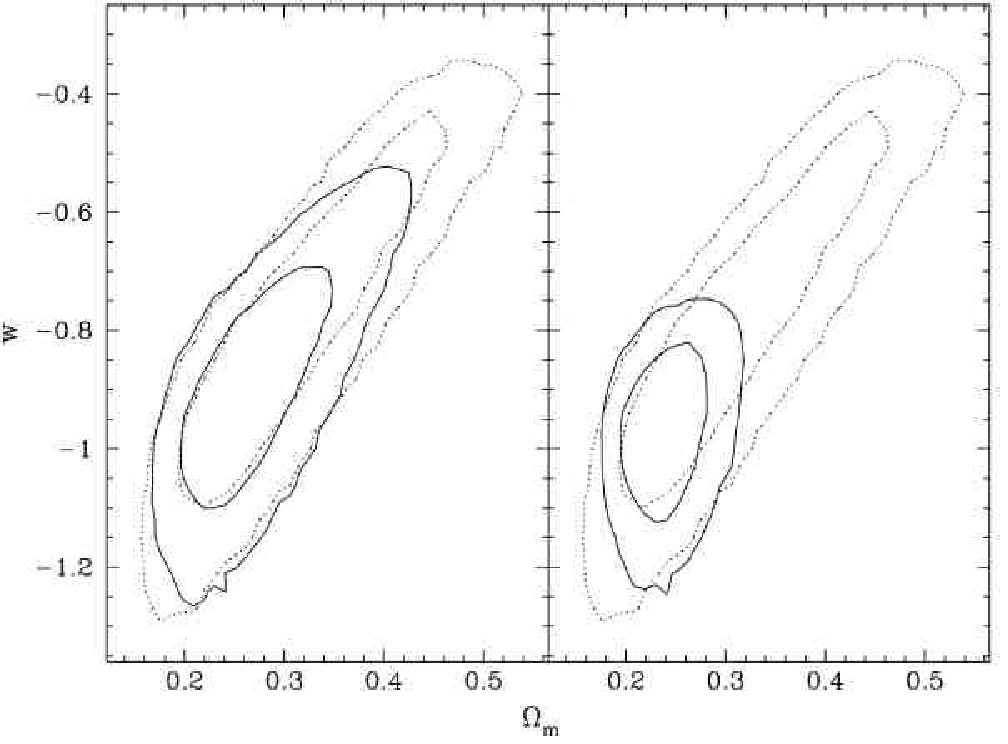}}
\caption{WMAP 3-year 1 and 2$\sigma$ error contours (dashed lines) in the $\Omega_M-w$~plane, for the {\tt wcdm+nopert} model of \citet{spergel06}. The solid contours represent the improvement obtained by using priors on $H_0$. Left panel: prior of $H_0=72\pm7$\ksm\ \citep{freedman01}. Right panel: prior of hypothetical future measurement of $H_0=74\pm3.5$\ksm.}
\label{fig:wmap}
\end{figure}

\vfill

\begin{figure}[ht]
\center{\includegraphics[scale=1.2]{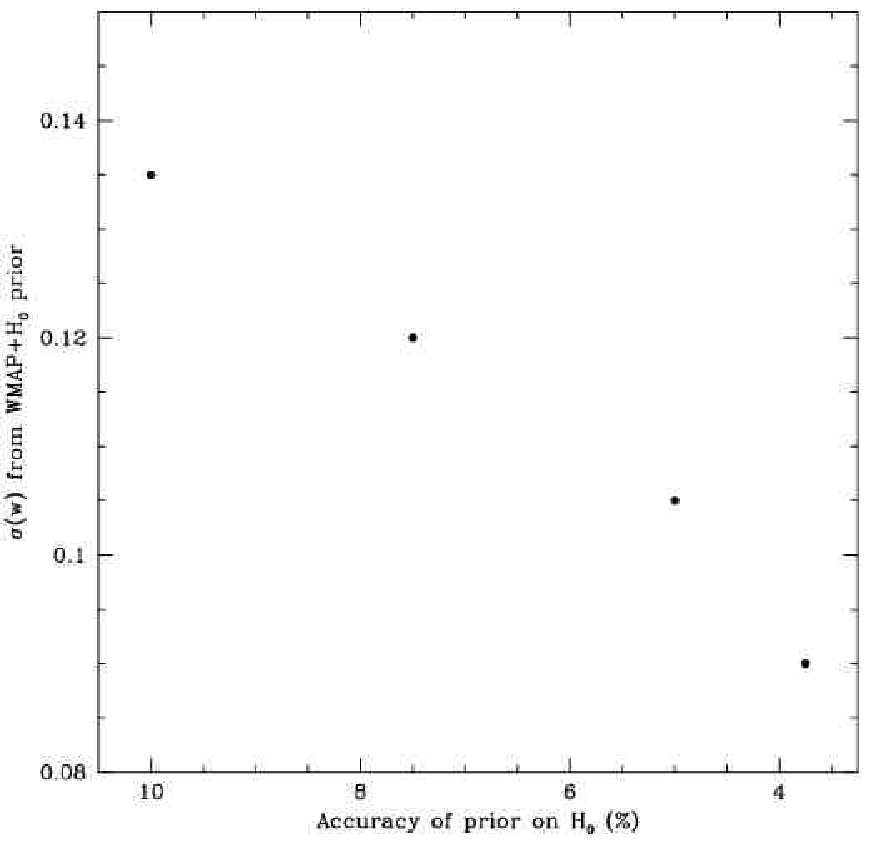}}
\caption{1$\sigma$ uncertainty in the value of $w$ for the {\tt wcdm+nopert} model of \citet{spergel06}, after including hypothetical priors on $H_0$ with a value of 74\ksm\ and decreasing uncertainty.}
\label{fig:h0w}
\end{figure}

\vfill\clearpage\ \par\vfill

\tabletypesize{\normalsize}
\tablewidth{0pt}
\begin{deluxetable}{llllllll}
\tablecaption{Log of Observations}
\tablehead{\colhead{Visit} & \colhead{UT Date} & \multicolumn{6}{c}{MJD at mid exposure$^*$} \\
\colhead{\#} & & \colhead{B} & \colhead{I} & \colhead{V} & \colhead{B} & \colhead{I} & \colhead{V}}
\startdata
O-01 & 2003 December 6  & 2980.0640 & 0.0734 & 0.0823 & 0.1206 & 0.1300 & 0.1389 \\
O-02 & 2003 December 7  & 2980.7980 & 0.8073 & 0.8163 & 0.8533 & 0.8627 & 0.8716 \\
\multicolumn{8}{l}{\it Table appears in its entirety in the full-resolution version of the paper}
\enddata
\label{tab:log}
\tablecomments{$*$: JD-2450000.0 for first exposure; thereafter, only the last five digits are given. $\dagger$: Guide star problems; limited usefulness. O: outer field; I: inner field.}
\end{deluxetable}

\vfill

\tabletypesize{\normalsize}
\tablewidth{0pt}
\begin{deluxetable}{llllllll}
\tablecaption{Secondary standards}
\tablehead{\colhead{ID} & \colhead{R.A.} & \colhead {Dec.} & \colhead{X} & \colhead{Y} & \colhead{V} & \colhead{I} & \colhead{B} \\
\colhead{} & \multicolumn{2}{c}{(J2000.0)}& \multicolumn{2}{c}{(pix)} & \multicolumn{3}{c}{(mag)}}
\startdata
O-42521 & 12:19:12.170 & 47:10:55.12 & 3587.1 & 4181.2 & 23.203(10) & 23.099(16) & 23.275(05) \\
O-42199 & 12:19:12.218 & 47:10:48.86 & 3704.5 & 4137.1 & 23.171(07) & 22.911(07) & 23.367(06) \\
\multicolumn{8}{l}{\it Table appears in its entirety in the full-resolution version of the paper}
\enddata
\label{tab:secstd}
\tablecomments{Uncertainties are given in parentheses and are expressed in units of $10^{-3}$~mag. O: outer field; I: inner field.}
\end{deluxetable}

\vfill

\tabletypesize{\normalsize}
\tablewidth{0pt}
\begin{deluxetable}{lrrcrr}
\tablecaption{Effects of selection criteria}
\tablehead{\multicolumn{1}{l}{Selection criteria (\S3.3)} & \multicolumn{2}{c}{$L_V > 0.75$} & & \multicolumn{2}{c}{$L_V > 2.0$}\\
\colhead{}& \colhead{Inner} & \colhead {Outer} & & \multicolumn{1}{c}{Inner} & \colhead {Outer}}
\startdata
Initial Sample                                        & 402 & 134 & & 195 &  63 \\
                                                      &     &     & &     &     \\
1. Amplitude ratios                                   &-110 & -28 & & -23 &  -4 \\
2. $\ebv < $~foreground ($2\sigma$)                   & -37 & -30 & & -17 & -20 \\
3. $\ebv > 0.5$~mag                                   &  -9 &  -1 & &  -6 &  -1 \\
4. $\Delta\mu_0 > 12$~mag \& $\sigma$ clipping        & -28 & -12 & & -11 &  -3 \\
                                                      &     &     & &     &     \\
{\bf Final sample}                                    & 218 &  63 & & 138 &  35 \\
\enddata
\label{tab:sel}
\tablecomments{This table shows the reduction in size of the different samples as a result of the selection criteria listed in \S3.3.}
\end{deluxetable}

\clearpage

\ \par\vfill

\tabletypesize{\scriptsize}
\tablewidth{0pt}
\begin{deluxetable}{lrrrrrrrrrrrr}
\tablecaption{{\normalsize Cepheid variables -- Basic data}}
\tablehead{\colhead{ID} & \colhead{Per.} & \colhead{R.A.} & \colhead{Dec.} & \colhead{X}   &  \colhead{Y} &\multicolumn{3}{c}{Magnitudes}            & \multicolumn{3}{c}{Amplitudes}       & \colhead{$L_V$} \\
                        & \colhead{(d)}  & \multicolumn{2}{c}{(J2000.0)}   & \multicolumn{2}{c}{(pixels)} &\colhead{V} & \colhead {I} & \colhead {B} & \colhead{V} & \colhead {I} & \colhead {B} &                   }
\startdata
 O-15165 &  3.30 & 12:19:20.668 & 47:10:31.19 & 3567.3 & 2383.1 & 26.646(042) & 26.092(040) & 27.148(054) & 503 & 168 & 691 & 2.01\\
 O-21998 &  3.36 & 12:19:17.808 & 47:10:11.81 & 4100.5 & 2836.1 & 26.540(043) & 25.959(043) & 26.993(061) & 482 & 214 & 812 & 2.26\\
\multicolumn{13}{l}{\it Table appears in its entirety in the full-resolution version of the paper}
\enddata
\label{tab:cephb}
\tablecomments{Errors in mean magnitudes are shown in parenthesis and expressed in units of $10^{-3}$~mag. Light curve semi-amplitudes are expressed in units of $10^{-3}$~mag. O: outer field; I: inner field.}
\end{deluxetable}

\vfill

\tabletypesize{\scriptsize}
\tablewidth{0pt}
\begin{deluxetable}{lrrrrrrrrrrrr}
\tablecaption{Rejected Cepheid candidates}
\tablehead{\colhead{ID} & \colhead{Per.} & \colhead{R.A.} & \colhead{Dec.} & \colhead{X}   &  \colhead{Y} &\multicolumn{3}{c}{Magnitudes}            & \multicolumn{3}{c}{Amplitudes}            & \colhead{$L_V$} \\
                        & \colhead{(d)}  & \multicolumn{2}{c}{(J2000.0)}   & \multicolumn{2}{c}{(pixels)} &\colhead{V} & \colhead {I} & \colhead {B} & \colhead{V} & \colhead {I} & \colhead {B} &                   }
\startdata
\multicolumn{13}{l}{\bf \boldmath $N_V < 18$ or $N_I < 18$ or $N_B < 18$}                                                         \\
 O-18637 &  3.21 & 12:19:21.128 & 47:11:27.81 & 2453.4 & 2607.1 & 27.000(051) & 26.281(052) & 27.409(077) & 361 & 190 & 474 & 1.27\\
 O-03121 &  3.23 & 12:19:27.280 & 47:10:20.74 & 3394.4 & 1030.0 & 27.158(030) & 26.623(113) & 26.981(027) & 359 &   0 &  29 & 0.92\\
\multicolumn{13}{l}{\it Table appears in its entirety in the full-resolution version of the paper}
\enddata
\tablecomments{Errors in mean magnitudes are shown in parenthesis and expressed in units of $10^{-3}$~mag. Light curve semi-amplitudes are expressed in units of $10^{-3}$~mag. O: outer field; I: inner field.}
\label{tab:rej}
\end{deluxetable}

\vfill

\tabletypesize{\normalsize}
\tablewidth{0pt}
\begin{deluxetable}{lrrrrrrr}
\tablecaption{Cepheid variables -- Derived properties}
\tablehead{\colhead{ID} & \colhead{Per.} & \colhead{$\mu_0^{VI}$} & \colhead {$\evi$} & \colhead {$\mu_0^{\rm av}$} & \colhead{$\ebv$} & \colhead{$r/$} \\
                        & \colhead{(d)}  & \colhead{(mag)}      & \colhead{(mag)}   & \colhead{(mag)}    &\colhead{(mag)}   & \colhead{$r_{iso}$}}
\startdata
 O-15165 &  3.30 & 11.009(115) & 0.022(025) & 11.009(075) & 0.016(025) & 1.30\\
 O-21998 &  3.36 & 10.900(122) & 0.022(025) & 10.901(075) & 0.016(025) & 1.23\\
\multicolumn{7}{l}{\it Table appears in its entirety in the full-resolution version of the paper}
\enddata
\tablecomments{$\mu_0^{VI}$: Extinction corrected distance modulus derived from $V$\&$I$ data (Eq.~\ref{eqn:muovi}). $\mu_0^{\rm av}$: Extinction corrected distance modulus derived from the average of Eqns.~(\ref{eqn:muovi}-\ref{eqn:muobv}). $\ebv$: Average value of extinction derived from all measured color excesses. Errors in distance moduli and extinction are shown in parenthesis and expressed in units of $10^{-3}$~mag. O: outer field; I: inner field.}
\label{tab:cephd}
\end{deluxetable}

\vfill

\clearpage\ \par\vfill

\tabletypesize{\normalsize}
\tablewidth{0pt}
\begin{deluxetable}{lccc}
\tablecaption{Error Budget of the Cepheid Distance Scale}
\tablehead{\multicolumn{1}{l}{Error source} & \colhead{Previous} & \colhead{This work} & \colhead{Goal}}
\startdata
{\bf A. Fiducial galaxy}   & {\bf LMC} &\multicolumn{2}{c}{\bf \gal}\\
S1. Distance modulus (sys) & 0.13 & 0.12 & 0.04   \\
R1. Distance modulus (ran) & \nd  & 0.09 & 0.02   \\
{\bf B. Photometric calibration}  &      &        \\
S2a. $V$ zeropoint         & 0.03 & 0.02 & 0.02   \\
S2b. $I$ zeropoint         & 0.03 & 0.02 & 0.02   \\
S2. Photometry (sys)       & 0.09 & 0.05 & 0.05   \\
R2. Photometry (ran)       & 0.05 & 0.03 & 0.02   \\
{\bf C. Extinction corrections}   &      &        \\
R3. Uncertainty in $R_V$   & 0.02 & 0.02 & 0.02   \\
R4. De-reddened PL fit     & 0.04 & 0.02 & 0.02   \\
{\bf D. Metallicity corrections}  &      &        \\
S3. Adopted correction     & 0.08 & 0.04 & 0.03   \\
\tableline
R$_T$. Total random        & 0.07 & 0.10 & 0.04   \\
S$_T$. Total systematic    & 0.18 & 0.14 & 0.07   \\
\tableline
{\bf Combined error (mag)} & {\bf 0.19} & {\bf 0.17} & {\bf 0.08}  \\
{\bf Combined error (\%) } & \ \ {\bf 10} & \ \ \ \ {\bf 8} & \ \ \ \ {\bf 4}  \\
\enddata
\label{tab:err}
\tablecomments{All errors expressed in magnitudes unless otherwise indicated. Previous: adapted from \citet{gibson00}. Goal: Anticipated reduction in uncertainties from \citet{humphreys07}, \citet{bersier07} and \citet{macri07a}.}
\end{deluxetable}

\vfill

\tabletypesize{\normalsize}
\tablewidth{0pt}
\begin{deluxetable}{llllllr}
\tablecaption{Updated distance moduli to high-quality type Ia SNe}
\tablehead{\colhead{Galaxy} & \colhead{SN} & \colhead{[O/H]} & \colhead{$\mu_0$} & \colhead{$\mu_{0,Z}$} & \colhead{$M^0_V$} & \multicolumn{1}{c}{Ref.}\\
\colhead{name} & & \multicolumn{1}{c}{(dex)} & \multicolumn{1}{c}{(mag)} & \multicolumn{1}{c}{(mag)} & \multicolumn{1}{c}{(mag)}  & }
\startdata
NGC$\,$3370   & 1994ae & $8.80\pm0.05$ & $32.23\pm0.04$ & $32.31\pm0.06$ & $-19.15\pm0.12$ & R05 \\
NGC$\,$3982   & 1998aq & $8.75\pm0.05$ & $31.56\pm0.08$ & $31.63\pm0.09$ & $-19.15\pm0.12$ & S01 \\
NGC$\,$4536   & 1981B  & $8.85\pm0.20$ & $30.80\pm0.04$ & $30.90\pm0.06$ & $-19.18\pm0.12$ & F01 \\
NGC$\,$4639   & 1990N  & $9.00\pm0.20$ & $31.61\pm0.08$ & $31.75\pm0.09$ & $-19.08\pm0.12$ & F01 \\
              &        &               &               &                &                 &     \\
\tableline
              &        &               &               &                &                 &     \\
{\bf Average} &        &               &               & \multicolumn{3}{r}{\boldmath $M^0_V=-19.14\pm0.07$}\\
\enddata
\tablecomments{$\mu_0$: Published Cepheid distance moduli; $\mu_{0,Z}:$ Distance moduli corrected for metallicity and our determination of the distance to the LMC. References: F01 = \citet{freedman01}; R05 = \citet{riess05}; S01 = \citet{stetson01}.}
\label{tab:snedm}
\end{deluxetable}

\vfill\clearpage

\tabletypesize{\scriptsize}
\tablewidth{0pt}
\begin{deluxetable}{llllllllll}
\tablecaption{Cepheid photometry}
\tablehead{\colhead{Visit} & \multicolumn{1}{c}{V} & \multicolumn{1}{c}{I} & \multicolumn{1}{c}{B} & \multicolumn{1}{c}{V} & \multicolumn{1}{c}{I} & \multicolumn{1}{c}{B} & \multicolumn{1}{c}{V} & \multicolumn{1}{c}{I} & \multicolumn{1}{c}{B}}
\startdata
   & \mc{\bf O-15165 \boldmath $P=3.30$~d}      & \mc{\bf O-21998 \boldmath $P=3.36$~d}      & \mc{\bf O-10450 \boldmath $P=3.71$~d}     \\
 1 & 26.900(096)  & 26.247(165)  & 27.532(121)  & 26.878(183)  & 25.915(120)  & 27.575(117)  & 26.565(116)  & 25.775(154)  & 26.898(109) \\
 2 & 26.797(163)  & 26.644(287)  & 26.448(325)  & 27.048(150)  & 26.091(208)  &     \nd      & 26.749(151)  & 26.068(123)  & 27.192(118) \\
 3 & 27.160(118)  & 26.413(162)  & 27.715(199)  & 27.090(160)  & 26.795(260)  & 27.453(272)  & 26.981(232)  & 26.310(219)  & 27.634(181) \\
 4 & 27.000(188)  & 26.367(161)  & 27.304(157)  & 27.178(207)  & 26.153(187)  & 27.661(365)  & 26.955(160)  & 26.114(163)  & 27.653(244) \\
 5 & 26.118(073)  & 25.837(117)  & 26.391(076)  & 25.992(062)  & 25.825(114)  & 26.137(068)  & 26.912(091)  & 26.472(214)  & 27.912(236) \\
 6 & 26.121(084)  & 25.973(144)  & 26.370(146)  & 25.942(089)  & 25.873(229)  & 26.199(104)  & 27.345(194)  & 25.990(117)  & 27.864(277) \\
 7 & 27.430(268)  &     \nd      & 27.672(170)  & 26.683(115)  & 26.082(126)  & 27.116(139)  & 26.127(088)  & 25.901(191)  & 26.461(093) \\
 8 & 26.994(198)  & 25.925(142)  & 27.453(167)  & 26.887(169)  & 26.252(180)  & 27.225(180)  & 26.268(083)  & 25.710(155)  & 26.591(144) \\
 9 & 26.283(084)  & 25.768(083)  & 26.304(082)  & 26.033(098)  & 25.567(120)  & 26.242(072)  & 26.607(127)  & 26.262(250)  & 27.656(200) \\
10 & 25.882(135)  & 25.868(165)  & 26.605(093)  & 26.074(097)  & 25.677(095)  & 26.222(064)  & 27.255(166)  & 25.975(155)  & 27.197(212) \\
11 & 27.057(247)  & 26.516(307)  & 27.971(295)  & 27.290(340)  & 26.589(342)  & 28.269(270)  & 26.768(186)  & 25.892(163)  & 26.691(078) \\
12 & 27.188(139)  & 26.129(103)  & 27.722(315)  & 26.773(105)  & 26.177(132)  & 27.532(207)  & 26.223(044)  & 26.346(206)  & 26.622(091) \\
13 & 26.727(171)  & 26.342(221)  & 27.554(137)  & 26.749(111)  & 25.874(165)  & 27.785(287)  & 27.022(256)  & 26.507(205)  &     \nd     \\
14 & 26.691(188)  & 25.951(117)  & 28.008(229)  & 26.812(142)  & 26.442(275)  & 27.492(234)  & 26.349(072)  & 25.814(135)  & 26.683(122) \\
\enddata
\label{tab:cepphot}
\tablecomments{The Julian Date for each visit can be found in Table~\ref{tab:log}. (*): Measurement deviated by more than $3\sigma$ from best-fit template light curve and was rejected. This table is available in its entirety upon request.}
\end{deluxetable}
\end{document}